\documentclass[aps,prx,twocolumn,superscriptaddress]{revtex4-2}
\usepackage{amssymb}
\usepackage{amsmath}
\usepackage{bm}
\usepackage{color}
\usepackage{float}
\usepackage{tikz}
\usepackage{graphicx}
\usepackage{hyperref} 
\usepackage{pgfplots,pgfplotstable}
\usetikzlibrary{positioning,decorations.markings,arrows.meta}
\usepackage{newtxtext,newtxmath} 
\usepackage[titletoc]{appendix}

\begin{document}


\title{A Universal Framework for Quantum Dissipation: \\Minimally Extended State Space and Exact Time-Local Dynamics}

\author{Meng Xu}
\affiliation{
Institute  for Complex Quantum Systems and IQST, Ulm University - Albert-Einstein-Allee 11, D-89069  Ulm, Germany}

\author{Vasilii Vadimov}
\affiliation{
QCD Labs, QTF Centre of Excellence, Department of Applied Physics, Aalto University, P.O. Box 15100, FI-00076 Aalto, Finland}

\author{Malte Krug}
\affiliation{
Institute  for Complex Quantum Systems and IQST, Ulm University - Albert-Einstein-Allee 11, D-89069  Ulm, Germany}

\author{J. T. Stockburger}
\affiliation{
Institute  for Complex Quantum Systems and IQST, Ulm University - Albert-Einstein-Allee 11, D-89069  Ulm, Germany}

\author{J. Ankerhold}
\affiliation{
Institute  for Complex Quantum Systems and IQST, Ulm University - Albert-Einstein-Allee 11, D-89069  Ulm, Germany}

\date{\today}

\begin{abstract}
The dynamics of open quantum systems is formulated in a minimally extended state space comprising the degrees of freedom of a system of interest and a finite set of non-unitary, pure-state reservoir modes. This formal structure, derived from the Feynman-Vernon path integral for the reduced density, is shown to lead to an exact time-local evolution equation in a mixed Liouville-Fock space. The crucial ingredient is a mathematically consistent decomposition of the reservoir auto-correlation in terms of harmonic modes with complex-valued frequencies and amplitudes, which are obtained from any given spectral noise power of the physical reservoir. This formulation provides a universal framework to obtain a family of equivalent representations which are directly related to new and established schemes for efficient numerical simulations. By restricting some of the complex-valued mode parameters and performing linear transformations, we make connections to previous approaches, whose auxiliary degrees of freedom are thus revealed as restricted versions of the minimally extended state space presented here. From a practical perspective, the new framework offers a computational tool which combines numerical efficiency and accuracy with long time stability and broad applicability over the whole temperature range and also for strongly structured reservoir mode densities. It can thus deliver high precision data with modest computational resources and simulation times for actual quantum technological devices.
\end{abstract}

\maketitle
\newpage

\section{Introduction}
All real-world quantum systems are in contact with agents which typically appear as reservoirs with many, very often macroscopically many, degrees of freedom \cite{gardiner2004quantum,breuer02,weiss12}. Open quantum systems exchanging energy and/or particles with these reservoirs thus constitute the generic setting encountered in experiments and to be described by theory. Ubiquitous is the situation, where a dedicated quantum system interacts with thermal reservoirs, giving rise to a wealth of phenomena in basically all fields of science from atomic physics and quantum optics \cite{gardiner2004quantum,scully1997quantum} to condensed matter physics \cite{kamenev2023field}, quantum information science \cite{nielsen2000quantum,averin2012macroscopic,beige2000quantum,verstraete2009quantum}, chemical physics \cite{nitzan06,may11}, and biology \cite{mohseni2014quantum}. However, apart from the fundamental interest in decoherence, dissipation induced phase transitions and phenomena alike \cite{weiss12}, in recent years, there has been a new quest for developing reliable and accurate descriptions of the dynamics of open quantum systems triggered by the impressive progress in quantum technologies \cite{georgescu202025,preskill2018quantum,cerezo2021variational,harrington2022engineered,reiter2017dissipative,Martin2022dissipative,theodoros2022criticality,ronzani2018tunable,hernandez2022autonomous,bouton2021quantum,opatrny2023nonlinear,david2023evolution,paladino2014noise}. 

\begin{figure*}[!ht]
\centering
\vspace*{-3cm}
\includegraphics[width=1.2\textwidth, keepaspectratio]{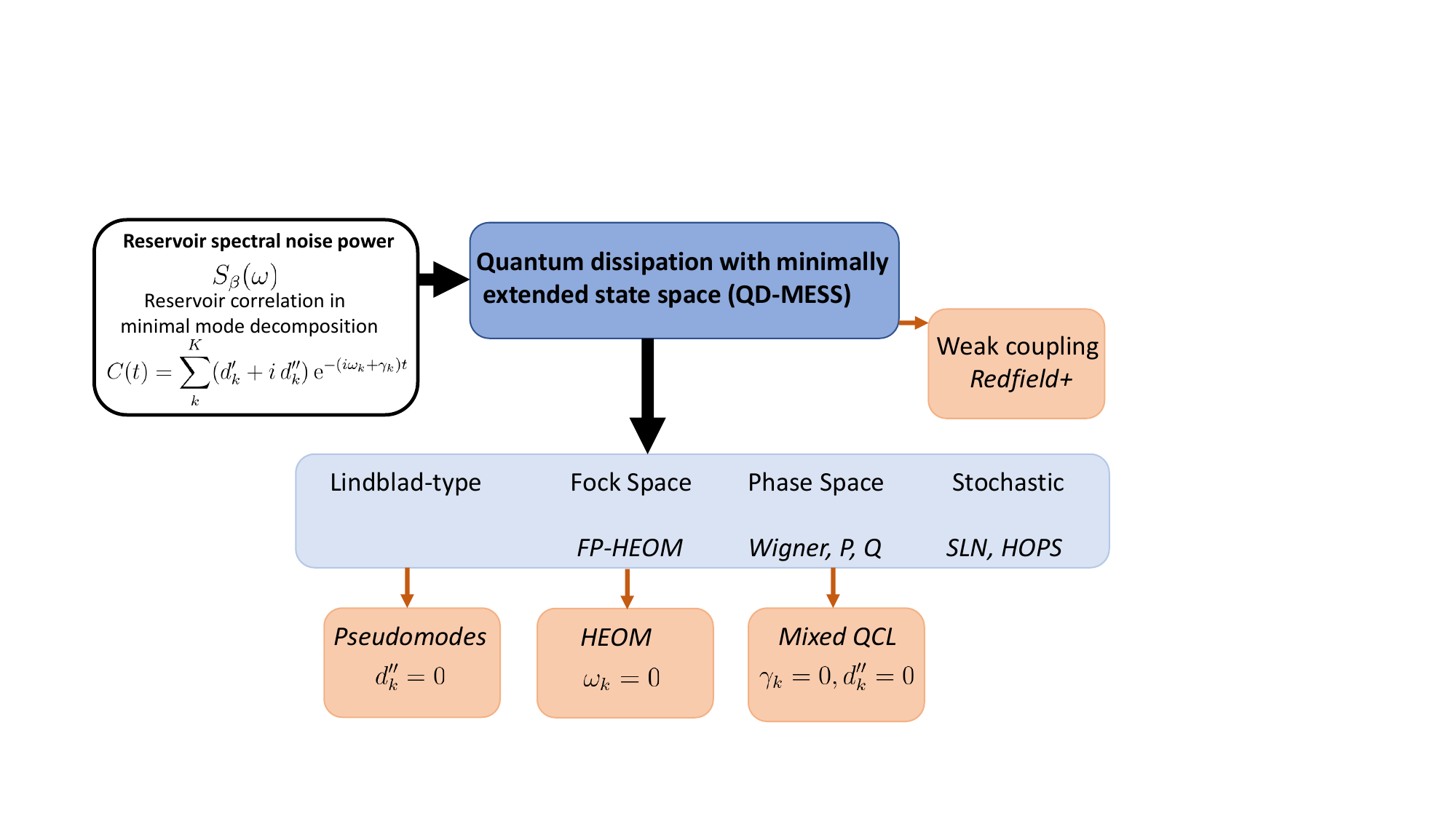}
\vspace*{-1.5cm}
\caption{Overview of the framework based on exact time-local equations for quantum dissipation with minimal state space (QD-MESS) developed in this work. The central ingredient is the spectral noise power $S_\beta(\omega)$ of a given thermal reservoir which serves as an input to obtain the decomposition of the reservoir correlation $C(t)$ in the time domain in minimal mode representation (white box upper left). This decomposition allows within the Feynman-Vernon path integral formulation to derive an exact time-local evolution equation in Liouville-Fock space (dark blue box). Several equivalent representations of this equation can be obtained (light blue box), mutually connected by similarity transformations, which are starting points for numerical simulation schemes. These include representations in Fock space (FP-HEOM), of Lindblad-type, in phase-space (for Wigner, P-, and Q distributions), and with stochastic fields (Stochastic Liouville-von Neumann/SLN and Hierarchy of Pure States/HOPS equations). Alternative/approximate treatments arise by simplifying the mode decomposition of $C(t)$ as indicated and for weak coupling (orange boxes).} 
\label{fig1}  
\end{figure*}

In fact, designing future quantum computing platforms and advanced quantum sensing protocols necessitates precise theoretical simulation schemes which go beyond established perturbative treatments \cite{breuer16,vega17,vega2008matter,groeblacher2015observation,gonzalez2019engineering,andersson2019non,giuseppe2019exciting,cygorek2022simulation,xu2022taming,nicholas2023effective}, for example, to also cover retardation effects (non-Markovianity) that inevitably appear at lower temperatures \cite{breuer02,weiss12}. The same is true for unconventional spectral reservoir properties with strongly structured mode densities such as bandgap materials introduced to provide protection against the detrimental influence of decoherence \cite{yousefi2022quantum,vats1998non,de2005two,bellomo2008entanglement,yang2013preservation} or to intentionally guide the system dynamics towards desired ground states \cite{liu2017quantum}. 

The general theoretical framework for open quantum systems in presence of thermal reservoirs is well-established based on system plus bath models and invoking projection operator techniques or path integral formulations \cite{weiss12,breuer02}. However, how to turn formally exact representations for the dynamics of reduced density operators into computationally accurate, efficient, and reliable simulation schemes are a very different story. A zoo of approaches has been developed in the past which can roughly be sub-grouped into three categories: Those that work directly with reduced densities, those that work in full Hilbert space of system+reservoir and those that, as a sort of compromise, work with hybrid schemes by properly 'unraveling' the intricate time-retardation imprinted by the reservoir onto the system dynamics.  While for a more detailed overview, we refer to the literature, here, only typical examples for each of these groups are mentioned: Path Integral Monte Carlo (PIMC) techniques stochastically sampling the Feynman-Vernon path integral have arguably the longest history with the main drawback being a degrading signal to noise ratio with increasing simulation time \cite{suzuki1993quantum,egger94,egger2000path,muhlbacher2005nonequilibrium}. As alternative path integral-based approaches we mention e.g.\ \cite{makri1995tensor,gull11,strathearn2018efficient, fux2021efficient,pollock2018non,bose2022multisite}. In full Hilbert space the Multi-Layer Multi-Configuration Time-Dependent Hartree (ML-MCTDH) \cite{beck00,wang2003multilayer} has been used to provide, for example, benchmark data for the paradigmatic spin-boson model \cite{wang10from} but becomes increasingly demanding for larger system sizes and longer simulation times; the same is true for related techniques, e.g.\ \cite{bulla03,vojta2005quantum,anders2007equilibrium,wu2013dynamics,werther2021coherent,ren2022time}. 
With respect to hybrid approaches, one concept is to introduce classical stochastic auxiliary fields to unravel the time-retarded reservoir correlation function in the influence functional of the Feynman-Vernon path integral to arrive at the Stochastic Liouville-von Neumann Equation (SLN) \cite{stockburger02} and the Stochastic Schrödinger Equation (SSE) \cite{diosi97}. An alternative methodology introduces auxiliary density operators (ADOs) instead to effectively unravel the time-retardation in a nested deterministic Hierarchical Equations of Motions (HEOM) \cite{tanimura89}, thus bypassing the need to average over large sets of noise samples. 

In parallel to these developments a variety of approximate treatments for open quantum systems have been brought to life  with limited applicability to certain ranges of parameters space such as weak coupling, elevated temperature, strong dissipation etc., but often with convincing performance. Examples include various versions of Born-Markov Master and Lindblad equations, the Redfield equation, mixed quantum-classical methods, and the Quantum Smoluchowski Equation \cite{breuer02, gardiner2004quantum,weiss12,risken84}.

The bottom line of the progress that has been made in the last years is that in terms of broad applicability, numerical stability, and computational efficiency hybrid treatments have gained some superiority against their extreme cousins. One may then pose the question whether the various approaches are actually family members growing out of a common theoretical framework. The first goal of this paper is to provide a positive answer to this question through the use of similarity transformations. The second goal, and a prerequisite of the first one, is to demonstrate that a uniform framework for quantum dissipation applicable over the whole temperature range and for arbitrary bath densities can be derived from the {\em time-nonlocal} Feynman-Vernon path integral in form of {\em time-local} quantum dissipation with a minmally extended state space (QD-MESS, see below). The latter comprises in addition to the system degrees of freedom a minimal (finite) set of effective reservoir modes (see Fig.~\ref{fig1}). 

More specifically, this common formulation appears when the quantum auto-correlation function of the reservoir is, in a mathematically consistent way and with arbitrary precision, decomposed  in a {\em finite} set of effective harmonic excitations with {\em complex-valued frequencies} $z_k=\gamma_k+i\omega_k, k=1, \ldots K$ and {\em complex-valued amplitudes} $d_k=d_k'+i d_k''$. Such a decomposition has been found recently to generalize the conventional HEOM to the Free-Pole-HEOM as a very efficient computational tool capturing the open system dynamics up to asymptotic times over the whole temperature range down to zero temperature and for strongly structured spectral bath densities \cite{xu2022taming,dan2023efficient}. 
The underlying physical picture, however, beyond a mere technicality remained unclear. Here, we reveal that, in fact, the FP-HEOM is nothing else than the Fock space representation of the QD-MESS. Alternative hybrid approaches such as a general pseudomode formulation, representations in phase space (Wigner, P, Q-representation), and stochastic unravelings (SLN and HOPS) follow from proper 'rotations' in Fock space preserving the commutator algebra. And there is even more to say: Namely, 
a comprehensive analysis of the QD-MESS elucidates that these exact representations reduce to various known approximate treatments when putting either of the four mode parameters in $z_k$ and $d_k$ to zero, see Fig.~\ref{fig1}. Eventually, the new QD-MESS offers a systematic perturbative expansion for weak coupling avoiding the commonly used Markov approximation.

The paper is organized as follows: We start in Sec.~\ref{sec:section2} to introduce the modeling of open quantum systems and the path integral formulation which only requires the spectral noise power as central ingredient to characterize reservoir properties. The central part is Sec.~\ref{sec:Liouville-Fock}, where we demonstrate how an optimal decomposition of the reservoir auto-correlations allows for an unraveling of the path integral expression through an 'un-reduction' of the density matrix imprinting the mode structure of the decomposition onto bosonic degrees of freedom. This in turn provides a time-local evolution equation for a quasi-density extended to a state in the resulting mixed Liouville-Fock product space. By representing this general equation in the number basis of the effective reservoir modes  one regains the known FP-HEOM equations. In Sec.~\ref{sec:section4} we show that the Liouville-Fock space equation is, in fact, a representative of an equivalent class of representations, mutually related to each other by similarity transformations. We particularly discuss the Lindblad-type and phase space representations from which also new and known simplified treatments emerge. Sec.~\ref{sec:section5} continues this analysis to stochastic schemes. The formulation of open quantum dynamics in the minimally extended state space allows to very efficiently implement matrix product states to boost its computational performance as presented in Sec.~\ref{sec:section6}. In addition, it provides the starting point for a systematic perturbative treatments in the weak coupling regime which avoids the often made Markov approximation Sec.~\ref{sec:perturbative}. Finally, main results are summarized and a Conclusions is given. 

\section{Modeling of dissipative quantum systems}\label{sec:section2}
To set the stage, we briefly recall the common modeling of open quantum systems based on system plus bath models \cite{weiss12}. The total Hamiltonian ($\hbar =k_B = 1$) is given by
\begin{equation}
    \hat H = \hat H_\mathrm s + \hat H_\mathrm {sb} + \hat H_\mathrm b \;\; ,
\end{equation}
where a system of interest, whose intrinsic properties are described by $\hat H_\mathrm s$, is embedded in a reservoir, typically a heat bath with bulk properties, i.e.,  many degrees of freedom, characterized by Hamiltonian $\hat H_\mathrm b$, governing free fluctuations, and system-bath interaction $\hat H_\mathrm{sb}$. Considering an initially factorizing state, the time-ordered and reservoir-averaged propagator of the system density matrix in the interaction representation is generally given by
\begin{eqnarray}\label{Eq:syspropgen}
\mathcal{J}(t)  &=& \langle \mathcal{U}(t)\rangle\nonumber\\
&=&\left\langle \mathcal{T} \exp\left(\int_0^t d\tau \mathcal{L}_{\rm sb}(\tau)\right) \right\rangle\, ,
\end{eqnarray}
where the superoperator $\mathcal{L}_{\rm sb} \cdot = - i [\hat H_{\rm sb},\cdot]$ has been introduced for the interaction. The angle brackets denote a reduction to superoperators in the system Liouville space, induced by an average with respect to the (initial) reservoir state $\rho_{\rm b}$ by a partial trace,
\begin{equation}
\mathcal{J}(t) \rho_{\rm s}(0) = 
\mathop{\mathrm{tr}_{\rm b}}\left\{
\mathcal{U}(t)
(\rho_{\rm s}(0)\otimes \rho_{\rm b})\right\},
\end{equation}
i.e., applying $\mathcal{J}$ to an initial system state $\rho_{\rm s}(0)$ yields the reduced density matrix at a later time  $\rho_{\rm s}(t)$.

Unless the dissipation is both intrinsically nonlinear and dominated by a small neighborhood of the system, the paradigm of a Gaussian environment applies. Assuming an unbiased interaction, the propagating superoperator $\mathcal{J}(t)$ is thus rendered as~\cite{kubo62,feynman63,aurel20}
\begin{equation}\label{Eq:sysproggauss}
\mathcal{J}(t) =  \mathcal{T} \exp\left(\frac{1}{2}\int_0^t ds\int_0^t d\tau \left\langle\mathcal{L}_{\rm sb}(s) \mathcal{L}_{\rm sb}(\tau) \right\rangle\right).
\end{equation}
Assuming a separable interaction, $\hat H_{\rm sb} = \hat{q}_{\rm s}\otimes\hat{X}_{\rm b}$, and with indices $\pm$ denoting left and right superoperators, this yields an operator equivalent of the Feynman-Vernon influence functional~\cite{feynman63}, i.e.\ $\mathcal{J}(t) = \mathcal{T} \exp({F}[\hat{q}_\mathrm s^+, \hat{q_\mathrm s}^-])$ with
\begin{equation}\label{Eq:IFsuperop}
\begin{split}
F[\hat{q}_\mathrm s^+, \hat{q}_\mathrm s^-]=&-\int_{0}^{t}ds \int_{0}^{s}d\tau\, \left[\hat{q}_\mathrm s^{+}(s)-\hat{q}_\mathrm s^{-}(s)\right] \\ 
  \times&\left[C(s-\tau)\hat{q}_\mathrm s^{+}(\tau) - C^{\ast}(s-\tau)\hat{q}_\mathrm s^{-}(\tau) \right]\, .
\end{split}
\end{equation}
Here $C(t)$ is the correlation function describing the free thermal fluctuations of the reservoir observable $\hat{X}_\mathrm b$,
\begin{equation}
C(t) = \mathop{\mathrm{tr}_{\rm b}} \left\{X_{\rm b}(t)X_{\rm b}(0)\rho_{\rm b}\right\} \;\;.
\end{equation}
Since thermal fluctuations follow the fluctuation-dissipation theorem, the reservoir's fluctuation spectrum $S_\beta(\omega)=\int dt\, {\rm e}^{i\omega t}\, C(t)$ can be obtained from the inverse reservoir temperature $\beta = 1/T$ and the dissipative response of the reservoir, i.e. 
\begin{equation}
S_\beta(\omega) = 2 n_\beta(\omega)\,  J(\omega)\,\, .
\end{equation}
Here, the spectral density $J(\omega)$ is introduced as an anti-symmetric function and $n_{\beta}(\omega) = 1/[1-\exp(-\beta\omega)]$ is the Bose distribution. In turn, the above relation leads in the time domain to
\begin{equation}\label{Eq:fluctuations_coth}
C(t) = \int_0^\infty d\omega \frac{J(\omega)}{\pi}
\left( \coth\frac{\beta\omega}{2} \cos(\omega t) - i \sin(\omega t) \right)\,\, .
\end{equation}
Historically, the reservoir model was often constructed based on elementary excitations such as microscopic bosonic modes, e.g., phonons or plasmons, or a phenomenological oscillator model.

A very convenient framework to formulate Gaussian quantum dissipation {\em non-perturbatively} is the path integral representation as pioneered by Feynman and Vernon \cite{feynman63} which since then has been extensively used \cite{weiss12}. In analogy to Eqs.\ (\ref{Eq:syspropgen})--(\ref{Eq:IFsuperop}), the reduced density operator $\hat{\rho}_\mathrm s(t)$ is expressed as a functional integral over paths supported by a Keldysh contour,
\begin{equation}\label{Eq:path-integral}
  \rho_\mathrm s^\pm(t)=\int\mathcal{D}[q_\mathrm s^+, q_\mathrm s^-] \, \mathcal{A}_\mathrm s[q_\mathrm s^+,q_\mathrm s^-] \, \mathrm{e}^{F[q_\mathrm s^+,q_\mathrm s^-]}\, \rho_\mathrm s^\pm(0) \;\;.
\end{equation}
Here,  the bare action factor $\mathcal{A}_\mathrm s[q_\mathrm s^+,q_\mathrm s^-] = \exp\{iS[q_\mathrm s^{+}]-iS[q_\mathrm s^{-}]\}$ captures the quantum dynamics in the absence of a bath with $S[q_\mathrm s^+]$,  $S[q_\mathrm s^{-}]$ the corresponding actions associated with forward and backward paths, respectively. Correspondingly, the operator-valued influence functional (\ref{Eq:IFsuperop}) turns into a c-number valued influence functional $F[q_\mathrm s^+, q_\mathrm s^-]$ by replacing operators $\hat{q}_\mathrm s^\pm$ by paths $q_\mathrm s^\pm(t)$. The endpoints of these paths appear explicitly when matrix elements (conventionally with respect to position) of the initial and final densities are considered. Here and in the sequel, we use a shorthand notation and indicate this dependence by superscripts $\pm$ of the respective densities $\langle \rho^{\pm}\rangle = \langle q^- |\rho|q^+\rangle$.

The above modeling has turned out to be so powerful that it is considered as the main pillar of the theory of quantum dissipation \cite{weiss12}. The only information required about thermal environments are the spectral bath densities $J(\omega)$ and temperature $k_{\rm B} T$. Knowledge about actual microscopic degrees of freedom is not necessary which implies wide applicability.

For practical purposes, two hybrid strategies have turned out to be particularly successful in the last years to cast this formulation into efficient simulation schemes. One way is to take one of the results (\ref{Eq:sysproggauss}) with (\ref{Eq:IFsuperop}) or (\ref{Eq:path-integral}) and perform all the steps backwards, substituting a Gaussian process with the same correlation function $C(t)$, but with variables taken not from a physical reservoir model, but a probability space~\cite{stockburger02}. The alternative route is to consider a properly 'un-reduced' description in a 
smaller, abstract quantum state space~\cite{mascherpa2020}. This procedure seems to be far more amenable to computation than a fully reduced description (\ref{Eq:path-integral}).

Following this concept, we will derive a time-local evolution equation with minimally extended state space below, termed Quantum Dissipation with Minimally Extended State Space (QD-MESS) which, in fact, provides a universal framework for  Gaussian dissipation to which a large class of alternative approaches encompassing also previous ones can be related.

\section{Quantum dissipation with minimally extended state space}
\label{sec:Liouville-Fock}

A notable feature of influence functionals is their ability to describe arbitrarily long-ranged temporal self-interactions of the system dynamics. The consequences are far-reaching: First, a general time-local equation of motion for $\hat{\rho}_\mathrm s(t)$ equivalent to the dynamics encoded in Eq. (\ref{Eq:path-integral}) is not known. Conventional time-local master equations following from a perturbative treatment, possess a limited scope of application, such as being restricted to sufficiently elevated temperatures and weak system-bath coupling. Even within these restrictions, the dissipative terms of quantum master equations may be inaccurate or extremely difficult to determine in the case of complex level structures or driven systems~\cite{stock17,vadimov2020validity,wu22}. Moreover, a direct numerical evaluation of the path integral expression, for example, in form of PIMC simulations, is highly demanding and becomes prohibitively expensive at long times due to oscillating integrands (dynamical sign problem) \cite{suzuki1993quantum,egger94,egger2000path,muhlbacher2005nonequilibrium}.

Hence, as already mentioned, a treatment based on a properly 'un-reduced' formulation has turned out to be particularly promising. A very powerful approach is the nested deterministic HEOM for auxiliary density operators (ADOs) \cite{tanimura89,tanimura06,tanimura2020numerically,yan2014theory,yan2016dissipation,wang2022quantum}. While the HEOM has been successfully applied to a variety of settings in recent years, its original version becomes increasingly expensive for lower temperatures or structured spectral bath densities. Namely, at lower temperatures, the decomposition of the correlation $C(t)$ in terms of decaying exponentials (see below) which is the central ingredient of the HEOM, requires  an increasing number of thermal frequencies (Matsubara frequencies) $\nu_n=2 n \pi/\hbar\beta$. This in turn implies a severe limitation of the HEOM outside elevated temperatures.
To address this problem, several strategies aimed at reducing the number of thermal frequency modes have been proposed \cite{xu05,hu11,duan17,erpenbeck2018extending,rahman2019chebyshev,cui2019highly,ikeda2020generalization,chen2022universal,ritschel2014analytic,hartmann2017exact,liu14,lambert2019modelling,alexander2022fingerprint}, providing to some extent routes to overcome some of these limitations. However, situations with strongly structured reservoirs and/or long-time simulations at ultra-low temperatures, where low-frequency reservoir modes govern the dynamics, remain a severe challenge.

It has recently been shown that this drawback can be cured by extending the HEOM to the so-called free-pole HEOM (FP-HEOM) \cite{xu2022taming}, where the impact of the reservoir is effectively described by a {\em minimal  set of exponentials with complex-valued amplitudes and frequencies}. This decomposition remains highly accurate and efficient even at $T=0$. Here, we will demonstrate that the introduction of ADOs is not merely a technicality introduced ad hoc, but rather emerges in an elegant, natural way whenever $C(t)$  is given in multi-exponential form with complex-valued frequencies and amplitudes.

\subsection{Unraveling of influence functionals using complex-valued auxiliary paths}
\label{sec:unraveling}

The time-nonlocality indicated by the double time integral in  Eqs.\ (\ref{Eq:IFsuperop}) and (\ref{Eq:path-integral}) can be unraveled through a Hubbard-Stratonovich transformation in path space, i.e., by defining a Gaussian auxiliary path variable, shifting its path, and uncompleting the square. In previous work~\cite{stockburger02,zhou2008solving,stockburger2016exact}, this approach was followed using two complex-valued auxiliary paths with a positive, normalized Gaussian measure directly reproducing the complex-valued $C(t)$.

In the present work, we follow a new way of unraveling the influence functional in terms of a path integral related to coherent states~\cite{negel88}.  As a first step, we recall that according to \cite{xu2022taming} the bath correlations  can be efficiently decomposed into a finite set of modes with arbitrary precision in a finite frequency range, i.e.,
 \begin{equation}\label{Eq:bcfd}
    C(t) = \sum_{k=1}^{K}\,d_k\,\mathrm{e}^{-z_k\,t} \;\; t\geq 0
\end{equation}
with complex-valued amplitudes $d_k$ and complex-valued frequencies $z_k = \gamma_k + i\omega_k$, $\gamma_k>0$. The maximal number of modes $K$ is
implicitly determined by setting a small threshold for the (maximum-norm) error of  $S_{\beta}(\omega)$  along the real frequency axis, resulting in $S_{\beta}(\omega)$ having $K$ poles $\{z_k\}$ and residues $\{d_k\}$ in the complex half-plane relevant for positive times~\cite{xu2022taming}.

Introducing the integral operator
\begin{equation}\label{Eq:convolution}
\mathcal{G}_k:~v(s) \mapsto \int_0^s \mathrm d\tau\;  \mathrm e^{-z_k(s-\tau)} v(\tau)
\end{equation}
and using angle brackets $\langle \cdot,\cdot\rangle$ to denote the natural inner product on $L_2[0,t]$,
\begin{equation}\label{Eq:ip}
    \langle u, v\rangle = \int_0^t \mathrm ds\, {u}^*(s)\, {v}(s),
\end{equation}
the corresponding decomposition of the influence functional reads
\begin{equation} \label{Eq:ifk}
    F[q_\mathrm s^+, q_\mathrm s^-] = \sum\limits_{k=1}^K \left( F_k[q_\mathrm s^+, q_\mathrm s^-] + \bar F_k[q_\mathrm s^+, q_\mathrm s^-]\right) \;\;,\\
\end{equation}
where 
\begin{subequations} \label{Eq:phik}
\begin{align}
    F_k[q_\mathrm s^+,q_\mathrm s^-]  =\left\langle -\sqrt{d_k^\ast}(q_\mathrm s^+ - q_\mathrm s^-), \sqrt{d_k}\mathcal{G}_k[q_\mathrm s^+] \right \rangle \\
   \bar F_k[q_\mathrm s^+, q_\mathrm s^-] = \left\langle -\sqrt{d_k}(q_\mathrm s^- - q_\mathrm s^+), \sqrt{d_k^\ast}\bar{\mathcal{G}}_k[q_\mathrm s^-]\right\rangle
\end{align}
\end{subequations}
and where $\bar{\mathcal{G}}_k$ is obtained from $\mathcal{G}_k$ by substituting $z_k^\ast$ for $z_k$.

We now treat the individual terms of Eq. (\ref{Eq:phik}) separately, using a single complex-valued auxiliary variable $\phi_k(t)$ with a normalized measure
\begin{equation} \label{Eq:coherent}
\mathcal D[\phi_k^\ast, \phi_k]\, {\rm e}^{-\langle \phi_k,D_k\phi_k\rangle}
\end{equation}
for each term~\footnote{If necessary, this measure includes an ``$\epsilon\to 0$'' regularization factor and a normalization allowing its (formal) interpretation as a coherent-state propagator~\cite{negel88}}. We will later make use of the fact that $\langle \phi_k,D_k\phi_k\rangle$ is virtually identical to the coherent-state action functional of a harmonic mode. For now, we observe that the differential operator
\begin{equation}
D_k = \partial_{\tau} +z_k
\end{equation}
is a weak inverse of the integral operator of $\mathcal{G}_k$, i.e., $D_k \mathcal{G}_k D_k = D_k$. Hence shifts of $\phi_k$ and $\phi^*_k$, followed by uncompleting the square can be used to construct each term $F_k[q_s^+,q_s^-]$ in Eq.\ (\ref{Eq:phik}) through the identity \cite{altland2010condensed}
\begin{multline}\label{Eq:gpi2}
\int \mathcal{D}[{\phi}_k^*,{\phi}_k]\, \exp\left\{-\langle\phi_k,D_k\phi_k\rangle + \langle w,\phi_k\rangle + \langle\phi_k,w'\rangle \right\}\\
= \exp\left(\left\langle w,\mathcal{G}_k w' \right\rangle\right) \;\;.
\end{multline}
Using a similar expression for the inverse $\bar{D}_k$ of $\bar{\mathcal{G}}_k$ and identifying $\langle w,w'\rangle$ with the r.h.s. of Eq.~(\ref{Eq:phik}), the reduced density matrix (\ref{Eq:path-integral}) can be represented through an extended path integral
\begin{multline}\label{Eq:fullaction}
  \rho_\mathrm s^\pm(t)  = \int \mathcal{D}[q_s^+,q_s^-] \prod_{k=1}^K\mathcal{D}[\phi_k^\ast,\phi_k;\psi_k^\ast,\psi_k]\,
    \mathcal{A}_\mathrm{s}[q_\mathrm{s}^+,q_\mathrm{s}^-]\\
    \times e^{i S_k[\phi_k,\phi_k^*,q_s^+,q_s^-]
    + i\bar{S}_k[\psi_k,\psi_k^*,q_s^+,q_s^-]}\rho_\mathrm{s}^\pm(0)
\end{multline}
with
\begin{align}\label{Eq:actionSk}
S_k[\phi_k,\phi_k^*,q_\mathrm s^+,q_\mathrm s^-] &= \int_0^t \mathrm d\tau\; (i\phi_k^*\partial_\tau \phi_k - H_k(\phi_k,\phi_k^*))\\
\bar{S}_k[\psi_k,\psi_k^*,q_\mathrm s^+,q_\mathrm s^-] &= \int_0^t \mathrm d\tau \;(i\psi_k^*\partial_\tau \psi_k -
\bar{H}_k(\psi_k,\psi_k^*))
\label{Eq:actionSkbar}
\end{align}
and
\begin{align}\label{Eq:CSH}
    H_k(\phi_k,\phi_k^*) &= -i z_k \phi_k^*\phi_k - i\sqrt{d_k}(q_\mathrm s^+ - q_\mathrm s^-)\phi_k
    + i\sqrt{d_k} q_\mathrm s^+\phi_k^*\\
    \bar{H}_k(\psi_k,\psi_k^*) &= -iz_k^* \psi_k^*\psi_k
    -i\sqrt{d_k^*}(q_\mathrm s^- - q_\mathrm s^+)\psi_k + i\sqrt{d_k^*}q_\mathrm s^-\psi_k^*.
    \label{Eq:CSbarH}
\end{align}
We have thus given the reduced density matrix a formal path-integral representation with fully {\em time-local} action, which arises naturally from the multi-exponential decomposition (\ref{Eq:bcfd}). The number of auxiliary paths is kept minimal through a numerically optimal construction of the decomposition (\ref{Eq:bcfd}). The complex weights of the additional paths, given through Eqs. (\ref{Eq:actionSk}--\ref{Eq:CSbarH}), properly identify them as coherent-state path integrals of bosonic modes, which are widely used in many-particle physics~\cite{negel88}. Moving from the symbolic representation of the action $S_k$ to its properly defined discrete version, one finds that this identification is for a coherent-state path integral with boundary conditions (cf. Appendix~\ref{sec:CSappendix})
\begin{equation}
\label{eq:initialcond}
    \phi_k(0) = \phi_k(t) = \psi_k(0) = \psi_k(t) = 0 \;\;,
\end{equation}
i.e., the dynamics of the auxiliary bosons formally begins and ends in a vacuum state in spite of the fact that the coefficients $d_k$ and complex rates $z_k$ may represent a finite-temperature reservoir.

\subsection{Dynamical states in a Liouville-Fock space}

The sum of all action terms in Eq. (\ref{Eq:fullaction}) corresponds to an extended dynamics which can be described conventionally, using linear operators as generators and states taken from a linear space. What is unconventional here is the appearance of action terms involving both a coupling between the path pair $(q_s^+,q_s^-)$, describing a mixed state, and coherent-state complex paths, describing pure states. Thus, the resulting state space in the Schr\"odinger picture of the dynamics described by Eq. (\ref{Eq:fullaction}) is neither a Liouville space (mixed states) nor a quantum mechanical Hilbert space (pure states). Since the path integral (\ref{Eq:fullaction}) has a forward-backward path structure for the system paths, but not for the coherent-state path variables, the corresponding quantum states form a product space where one factor is the quantum Liouville space of the system, the other the Fock space of a $2K$-mode harmonic system,
\begin{equation}
    \mathsf{\Gamma} = \mathbb{L}_\mathrm s \otimes \mathbb{F}_{2K}\;.
\end{equation}
Assigning raising and lowering operators $\hat{a}_k^\dagger$, $\hat{a}_k$, $\hat{b}_k^\dagger$ and $\hat{b}_k$ to the pure-state modes described by the coherent-state paths $\phi_k$, $\phi_k^\ast$, $\psi_k$ and $\psi_k^\ast$, the dynamics of an extended state $\hat{W} \in \mathsf{\Gamma}$
reads
\begin{equation}\label{Eq:standard-heom-liouville-fock}
\begin{split}
    \dot{\hat{W}}
=& -i\mathcal{L}_\mathrm{s} \hat{W}  + \sum_k
\left\{ \hat{\Gamma}_k\hat{W} + \sqrt{d_k}\hat{q}_\mathrm{s}\hat{a}_k^{\dagger}\hat{W}  \right. \\
 & \left.  + \sqrt{d_k^*} \hat{b}_k^{\dagger}\hat{W}\hat{q}_\mathrm{s} -  [\hat{q}_\mathrm{s}, (\sqrt{d_k} \hat{a}_k - \sqrt{d_k^*} \hat{b}_k)\hat{W}]\right\} \;\; .
\end{split}
\end{equation}
Here,  $\mathcal{L}_s$ denotes the Liouville operator of the bare system and we introduce a generator  \begin{equation}
\hat{\Gamma}_k = -z_k\, \hat{a}_k^{\dagger}\hat{a}_k - z_k^*\, \hat{b}_k^{\dagger}\hat{b}_k \;\;
\end{equation}
describing the dynamics of the unperturbed auxiliary bosons.

In order to make contact with the reduced dynamics originally considered, both the initial preparation and the procedure equivalent to the partial trace over the reservoir must be identified here. Having established the boundary conditions of the coherent-state path integral in (\ref{eq:initialcond}), we conclude that the factorizing initial condition corresponds to an initial condition with all auxiliary bosons in the vacuum state, and tracing out the real reservoir modes corresponds to formally projecting all auxiliary bosonic modes onto their (formal, not physical) vacuum state.

This final observation leads to a first major finding of this paper: 
Observing that the quantum numbers $\hat{m}_k = \hat{a}_k^{\dagger}\hat{a}_k$ and $\hat{n}_k = \hat{b}_k^{\dagger}\hat{b}_k$ can be identified with the components of the multi-index identifying used in FP-HEOM to label auxiliary states, we conclude that the FP-HEOM dynamics is fully equivalent to that described by the path integral (\ref{Eq:fullaction}): It is \emph{identical} to the mixed-space dynamics Eq.~(\ref{Eq:standard-heom-liouville-fock}), given in the number basis for the auxiliary bosons: Indeed, using the decomposition
\begin{equation} \label{Eq:f-expansion}
\hat{W}(t) = \sum_{\bf m,n}\,\hat{\rho}_{\bf m,n}(t) |\bm{m},\bm{n}\rangle \,
\end{equation}
with multi-indices $\{{\bf m,n}\} = \{m_1,n_1,\ldots,m_K,n_K\}$, we arrive at the balanced FP-HEOM equation used in~\cite{xu2022taming}, 
\begin{equation}\label{Eq:standard-heom}
\begin{split}
\dot{\hat{\rho}}_{\bf m,n} =
& -\left(i\mathcal{L}_s
+\sum_k m_{k} z_{k} + \sum_k n_{k} z_{k}^{*} \right) \hat{\rho}_{{\bf m,n}}  \\
& -\sum_k \sqrt{(m_k+1)\, d_k} \left[\hat{q}_\mathrm s,\hat{\rho}_{{\bf m}_k^{+},{\bf n}} \right] \\
& +\sum_k \sqrt{(n_k+1)\, d_k^*}\left[\hat{q}_\mathrm s,\hat{\rho}_{{\bf m,n}_k^{+}} \right]  \\
& +\sum_k \sqrt{m_k d_k}\, \hat{q}_\mathrm s \hat{\rho}_{{\bf m}_k^{-},{\bf n}} \\
& +\sum_k \sqrt{n_k d_k^*}\, \hat{\rho}_{{\bf m,n}_k^{-}}\hat{q}_\mathrm s
\;\;.
\end{split}
\end{equation}
Here a subscript $k$ accompanied by a raised ``$+$'' or ``$-$'' indicates a multi-index where the $k$-th element has been raised or lowered by one (relative to the multi-indices appearing in the left-hand side). Since no basis has been given for the system degrees of freedom, the projection
\begin{equation}
\hat{\rho}_{\bf m,n} = \langle{\bf m,n}|\hat{W}
\end{equation}
results in terms which still have the character of a density operator in the system Liouville space. As stated before in different language, the physical density, i.e.\, the reduced density operator, follows as
\begin{equation}
\label{eq:red-FP-HEOM}
\hat\rho_\mathrm s=\hat\rho_{\bf 0,0}\, ,
\end{equation}
while all other terms $\hat{\rho}_{\bf m,n}$ appearing in the dynamics are labeled \emph{auxiliary density operators} (ADOs).

The FP-HEOM thus appears in a natural, perhaps even cogent manner as a minimal un-reduction of the Feynman-Vernon expression for the reduced density matrix. The appearance of modes constrained to the ground state at times $0$ and $t$ may startle or give the appearance of a method only suitable for zero-temperature environments. 
However, the auxiliary bosons are not necessarily physical entities, but can be viewed as computational tools which include the proper relationship between positive- and negative-frequency parts of the reservoir power density given by the fluctuation-dissipation theorem by matching the coefficients $d_k$ and $z_k$ in Eq. (\ref{Eq:bcfd}) to any {\em physical} $C(t)$ at any temperature. Formally, {\em using pure-state auxiliary modes representing mixed reservoir states leads to a more compact computational state representation}.

Another major gain made in the present derivation of the more abstract form (\ref{Eq:standard-heom-liouville-fock}) of FP-HEOM \cite{xu2022taming} lies in the fact that the HEOM methodology \cite{tanimura89,tanimura06,tanimura2020numerically} is now no longer tied to the number presentation of the Fock space factor in the dynamical state space. Other representations, e.g., phase space representations, and useful transformations of the Fock space will be explored in later sections.

As a first example, a simple transformation will be given here: Merely declaring half of the auxiliary bosons as residing in a dual Fock space defines an alternative global state $\rho$ linked to the ADO hierarchy through
\begin{equation}
    \hat\rho = \sum_{\bf m,n}|{\bf m}\rangle\hat \rho_{\bf m,n}\langle {\bf n}|\,,\quad
    \hat \rho_{\bf m,n} = \langle{\bf m}|\hat\rho|{\bf n}\rangle\,.
\end{equation}
This ansatz creates the appearance of a Liouville-Liouville product space, with one factor spanned by the basis~$\{|{\bf m}\rangle\langle {\bf n}|\}$. Formally, one 
uses in Eq. (\ref{Eq:standard-heom-liouville-fock}) the correspondences
\begin{eqnarray}
    \hat{a}_k \hat{W} &\leftrightarrow& \hat{a}_k \hat{\rho},\quad
    \hat{a}_k^\dagger \hat{W} \leftrightarrow \hat{a}_k^\dagger \hat{\rho} \, \nonumber \\
    \hat{b}_k \hat{W} &\leftrightarrow&  \hat{\rho} \hat{a}_k^{\dagger}, \quad
    \hat{b}_k^\dagger \hat{W} \leftrightarrow\hat{\rho} \hat{a}_k\; \nonumber
\end{eqnarray}
 resulting in an equation of motion for $\hat\rho$, i.e.,  
\begin{multline}\label{Eq:standard-heom-density}
    \dot{\hat{\rho}} = -i [\hat{H}_s + \sum_k\hbar\omega_k \hat{a}_k^\dagger\hat{a}_k,\hat\rho]
     - \sum_k\gamma_k\{\hat{a}_k^\dagger\hat{a}_k,\, \hat{\rho}\} \\
    + \sum_k \left[\sqrt{d_k}(\hat{q}_s\hat{a}_k^\dagger\hat\rho - [\hat{q}_s,\hat{a}_k\hat\rho])
     +\mbox{h.c.}\right]
    \;\;,
\end{multline}
where the hermitian conjugate relates to the scalar product
\begin{equation}
    \langle \hat\sigma, \hat\rho\rangle = \sum_{\bf m,n} \mathop{\mathrm{tr}}\hat\sigma_{\bf n,m}^\dagger \hat\rho_{\bf m,n}\;.
\end{equation}
Equation (\ref{Eq:standard-heom-density}) reveals that an initially hermitian $\hat\rho$ will remain hermitian~\footnote{However, other properties of a true density matrix seem to be lacking.}. 
As previously, the free dynamics in the bosonic factor space (with $d_k\equiv 0$) has no immediate physical meaning by itself---there is damping, but no decoherence. However, the following sections will show that one can arrive at a different conclusion after performing suitable transformations.

 The  two expressions (\ref{Eq:standard-heom-liouville-fock}) and (\ref{Eq:standard-heom-density}) serve as the basis for a number of further developments that we will present in the sequel: First, they can be mapped onto representations in Lindblad- and phase-space form, second, they provide a direct link to known stochastic unraveling schemes, and third, they offer the starting point for perturbative treatments. Henceforth, we will refer to these two fundamental time-local evolution equations as Quantum Dissipation with Minimally Extended State Space, i.e.\ QD-MESS.

\section{Mapping to alternative representations}
\label{sec:section4}

The representation of open quantum dynamics in minimal state space according to Eq. (\ref{Eq:standard-heom-liouville-fock}) 
is, in fact, not unique. There is a whole class of equivalent representations which can be obtained via similarity transformations $\mathcal{S}$ in Fock space with Det$\{\mathcal{S}\}=1$ (unimodular transformations) so that the operator algebra is preserved. We mention in passing that the well-known Bogoliubov transformations also belong to this type of transformation. Here, we demonstrate the corresponding mapping onto two representations of particular relevance, namely, of Lindblad-type and in phase space (see Fig.~\ref{fig1}). 

\subsection{Lindblad structure}
We first turn to the Lindblad structure and conveniently start from Eq. (\ref{Eq:standard-heom-liouville-fock}) with the introduction of the following operator 
\begin{equation}\label{Eq:transformation}
    \mathcal{S} = \exp\left\{-\sum_k\hat{a}_k\hat{b}_k\right\}\exp\left\{\sum_k(\mu_k\hat{m}_k+\nu_k\hat{n}_k)\right\}\;\;.
\end{equation}
Here, parameters $\mu_k, \nu_k$ are determined via $e^{\mu_k} = i\sqrt{d_k/d_k'}$ and $e^{\nu_k} = -i\sqrt{d_k^\ast/d_k'}$, respectively. The first exponential mixes the modes $\hat{a}_k$ and $\hat{b}_k$, while the latter one scales the coefficients. The action of $\mathcal{S}$ on the creation-annihilation operators reads
\begin{subequations}\label{Eq:similarity-transformation}
\begin{align}
    \mathcal{S} \hat{a}_k^{\dagger} \mathcal{S}^{-1} &= {\rm e}^{\mu_k}\,(\hat{a}_k^{\dagger} -\hat{b}_k) \;\; \\
    \mathcal{S} \hat{b}_k^{\dagger} \mathcal{S}^{-1} &= {\rm e}^{\nu_k}\,(\hat{b}_k^{\dagger} - \hat{a}_k) \;\; \\
    \mathcal{S} \hat{a}_k \mathcal{S}^{-1} &= {\rm e}^{-\mu_k}\,\hat{a}_k \;\; \\
    \mathcal{S} \hat{b}_k \mathcal{S}^{-1} &= {\rm e}^{-\nu_k}\,\hat{b}_k \;\;.
\end{align}
\end{subequations}
This time-independent similarity transformation can be understood as a rotation in combination with a scaling operation which preserves the commutator relations.

Now, upon implementing the above transformation on Eq. (\ref{Eq:standard-heom-liouville-fock}), one arrives with $\hat{W}_{S} = \mathcal{S}\hat{W}$ in Liouville space at
 \begin{multline}\label{Eq:transformed-HEOM-1}
\dot{\hat{W}}_{\mathcal{S}} = -i\mathcal{L}_s \hat{W}_{\mathcal{S}} + \sum_k\left(2\gamma_k \hat{a}_k\hat{b}_k-z_k\hat{a}_k^\dagger\hat{a}_k - z_k^\ast \hat{b}_k^\dagger\hat{b}_k 
\right)\hat{W}_{\mathcal{S}} \\
+ \sum_k \left\{i\frac{d_k}{\sqrt{d_k'}}\hat{q}_s (\hat{a}_k^{\dagger} - \hat{b}_k) \hat{W}_{\mathcal{S}} - i\frac{d_k^\ast}{\sqrt{d_k'}} (\hat{b}_k^\dagger - \hat{a}_k) \hat{W}_{\mathcal{S}}\hat{q}_s \right. \\  \left.
+ i\sqrt{d_k'}[\hat{q}_s, (\hat{a}_k + \hat{b}_k)\hat{W}_{\mathcal{S}}] \right\}\;\;.
 \end{multline}
In parallel to Eq.~(\ref{Eq:standard-heom-density}), this expression can be mapped onto a time evolution equation for the density matrix which is almost in Lindblad form, i.e., 
\begin{multline}\label{Eq:standard-heom-lindblad}
    \dot{\hat{\rho}}_{\mathcal{S}} = -i[\hat{H}_{\rm eff},\hat{\rho}_{\mathcal{S}}] 
   +\sum_k \frac{d_k''}{\sqrt{d_k'}} ([\hat{a}_k, \hat{\rho}_{\mathcal{S}}]\hat{q}_\mathrm s - \hat{q}_\mathrm s [\hat{a}_k^\dagger, \hat{\rho}_{\mathcal{S}}]) \\
    + 2\sum_k \gamma_k [\hat{a}_k  \hat{\rho}_{\mathcal{S}} \hat{a}_k^{\dagger} -\frac{1}{2} \{\hat{a}_k ^{\dagger} \hat{a}_k, \hat{\rho}_{\mathcal{S}} \}] \;\; 
\end{multline}
with the effective Hamiltonian
\begin{equation}\label{Eq:heff}
    \hat{H}_{\rm eff}=\hat{H}_\mathrm s + \sum_k[\omega_k\hat{a}_k^\dagger\hat{a}_k - \sqrt{d_k'}\hat{q}_\mathrm s(\hat{a}_k^\dagger+\hat{a}_k) ]\,\, .
\end{equation}
One observes that the right hand side is hermitian with a vanishing trace. Accordingly, the corresponding time evolution is norm conserving. However, while the dissipator (last term) is in Lindblad form, the $d_k''$-dependent coupling part is not of Hamiltonian form. Positivity of the density operator in extended space is thus not guaranteed. 

Due to the state basis having been modified under the action of the operator $\mathcal{S}$ (\ref{Eq:transformation}), the boundary conditions for the auxiliary modes are now this:  The initial boson state for a factorizing initial condition is still the vacuum, but the final state to give the reduced density $\hat{\rho}_s$ is no longer constrained to it and can be calculated from
\begin{align}\label{Eq:recal-rho}
    \hat{\rho}_\mathrm s(t) &= \langle {\bf 0,0}|\mathcal{S}^{-1} W_S(t) \rangle \nonumber \\
    & = \sum_{\bf n} \hat{\rho}^{(\mathcal{S})}_{\bf n,n}(t)  = \mathrm{Tr_B} \hat{\rho}_{\cal S}(t)\;\;.
\end{align}
The above relation suggests that the combined trace of the auxiliary bosonic modes and the system degrees is conserved and equals the identity. However, the new reduced density calculation incorporates, in contrast to (\ref{eq:red-FP-HEOM}),  excited states, thereby making the calculation of steady states more expensive compared to the counterpart in (\ref{Eq:standard-heom-liouville-fock}).

\begin{figure}[!h]
\centering
\resizebox{0.45\textwidth}{!}{\begin{tikzpicture}
   \begin{axis}[
        width=\linewidth,
        height=0.8\linewidth,
        xlabel={$\omega_\epsilon$},
        ylabel={K},
        label style={font=\footnotesize\Large},
        tick label style={font=\footnotesize\Large},
        xmin=1e-7, xmax=1e-1,
        ymin=15, ymax=50,
        line width=1pt,
        tickwidth=1pt,
        major tick length=4pt,
        grid=major,
        grid style={gray!10, line width=0.5pt, dashed},
        major tick style={black, line width=1pt},
        minor tick style={black, line width=0.5pt},
        axis line style={line width=1pt},
        xmode=log,
    ]
        \addplot[
            black,
            only marks,
            mark=*,
            mark options={scale=0.8, fill=black},
        ] table {
x    y
1.00E-07	50
1.00E-06	47
1.00E-05	38
1.00E-04	30
1.00E-03	26
1.00E-02	22
1.00E-01	18
        };

        \addplot[
            no markers,
            red,
            smooth,
        ] table[
            y={create col/linear regression={y=y}},
        ]{
x    y
1.00E-07	50
1.00E-06	47
1.00E-05	38
1.00E-04	30
1.00E-03	26
1.00E-02	22
1.00E-01	18
        };

    \end{axis}
\end{tikzpicture}}
\vspace*{-0.5cm}
\caption{Number of auxiliary modes $K$ required to represent $S_\beta(\omega)$ in the frequency domain $\mathcal{A} = [-\omega_D,-\omega_\epsilon] \cup [\omega_\epsilon, \omega_D]$ for a sub-ohmic spectral density $J(\omega)=\alpha \omega_c^{1-\eta}\omega^\eta {\rm e}^{-\omega/\omega_c}$. Parameters are $\eta= 1/2$, $\alpha = 0.05$, $\omega_c = 10$, $T =0$, $\omega_D = 10^2$, the accuracy is chosen as $\delta = 10^{-9}$, in arbitrary unit. The logarithmic increase of reservoir modes for long times, i.e. $K\propto -\ln(\omega_\epsilon)\propto \ln(t_{\rm max})$ with $t_{\rm max}$ being the maximal simulation time range, is in stark contrast to the polynomial growth for  alternative approaches
\cite{trivedi2021convergence,trivedi2022description}, see text for details.}
\label{fig2}
\end{figure}
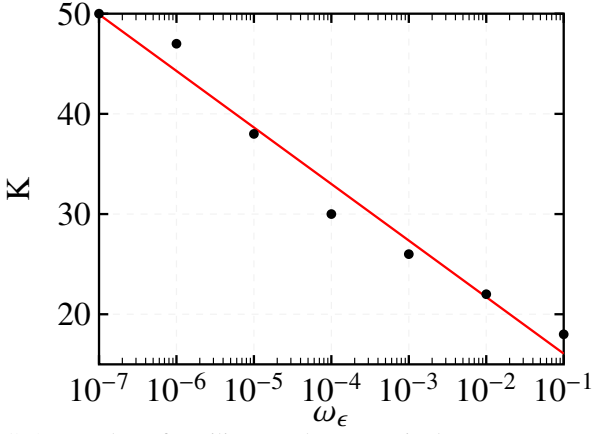

In recent years, a number of schemes for simulating dynamically quantum dissipation have been proposed \cite{pleasance2020generalized,pleasance2021pseudomode,tamascelli2018nonperturbative,mascherpa2020,arrigoni2013nonequilibrium,trivedi2021convergence,tamascelli2019efficient,lentrodt2020ab,nusseler2022fingerprint,medina2021few,cirio2022pseudo,sanchez2022few}, which start with a reservoir correlation decomposition similar to that in Eq. (\ref{Eq:bcfd}) and formulate the time evolution with equations in complete Lindblad form. Here, as seen above, we only regain such a form by restricting the amplitudes to real values $d_k\equiv d_k', d_k''=0$, in (\ref{Eq:standard-heom-lindblad}), see also Fig.~\ref{fig1}. We note that if $d_k \in \mathbb{R}^+$ and $\gamma_k = 0$, the dynamics in extended space in (\ref{Eq:standard-heom-lindblad}) is unitary while the reduced is not. However, constraining the decomposition to real-valued amplitudes $d_k'$ \cite{mark1}, comes with a significant limitation in that the number of auxiliary modes (then termed pseudomodes) escalates significantly, approximately polynomially \cite{trivedi2021convergence,trivedi2022description}, at lower temperatures and for more intricate spectral bath densities. Conversely, for the QD-MESS, the growth in the number of auxiliary modes is more favorable as it only increases logarithmically, see Fig.~\ref{fig2}, thereby enabling simulations to be conducted up to asymptotic times \cite{xu2022taming}.

\subsection{Phase space representation}

In quantum optics, several representations have been developed to write down phase space distributions. 
The most prominent ones are the Glauber-Sudarshan $P$ representation, the $Q$ representation, and Wigner $W_W$ representation \cite{gardiner2004quantum}. 

Here, based on (\ref{Eq:standard-heom-lindblad}) we derive corresponding results for distributions of auxiliary bosonic modes while keeping the density operator in the system degree of freedom. Accordingly, the auxiliary modes are transformed via 
\begin{subequations}\label{Eq:phasespace}
\begin{align}
a_k \hat{\rho}_{\cal S} \leftrightarrow \left(\alpha_k-\frac{\sigma-1}{2}\frac{\partial}{\partial\alpha_k^*}\right) \Phi_\sigma \;\;\,\\
 a_k^\dagger \hat{\rho}_{\cal S} \leftrightarrow  \left(\alpha_k^*-\frac{\sigma+1}{2} \frac{\partial}{\partial \alpha_k}\right) \Phi_\sigma \;\;\,
 \end{align}
 \end{subequations}
with a parameter $\sigma = 1, -1, 0$ according to system operator-valued distributions $\Phi_1\equiv P$, $\Phi_{-1}\equiv Q$, and $\Phi_0\equiv W_W$ in $(\{\alpha_k, \alpha_k^*\})$-FP-mode space.  Together with corresponding relations for the right operators, this set of mapping rules (\ref{Eq:phasespace}) defines again an unimodular similarity transformation, thus preserving commutation relations.

As a consequence, the new global state $\Phi_\sigma$ is a function of coherent-state labels instead of multi-indices. It is still an operator residing in the Liouville space of the system, like the reduced density matrix and the ADOs employed in HEOM. Thus, we arrive at the  following Liouville Fokker-Planck equations for the respective phase space distributions
\begin{eqnarray}
\dot{\Phi}_\sigma(t) &=& \Big[\mathcal{L}_\sigma + \sum_k\left(z_k\partial_{\alpha_k}\alpha_k + z_k^\ast\partial_{\alpha_k^\ast} \alpha_k^\ast\right)\nonumber\\
&& + (1-\sigma)\sum_k \gamma_k\partial^2_{\alpha_k\alpha_k^\ast}\Big]\  \Phi_\sigma(t) \;\;
\end{eqnarray}
with the enlarged super-operator $\mathcal{L}_\sigma \Phi_\sigma = -i[H_\mathrm s,\Phi_\sigma]+\sum_k \mathcal{D}_k^{(\sigma)} \Phi_\sigma$, where
\begin{multline}
    \mathcal{D}_k^{(\sigma)}\Phi_\sigma = - i\sigma\frac{\sqrt{d_k'}}{2}  (\partial_{\alpha_k^\ast} + \partial_{\alpha_k}) [\hat{q}_\mathrm s,\Phi_\sigma ]  \\
    -i\frac{\sqrt{d_k'}}{2}  (\partial_{\alpha_k} - \partial_{\alpha_k^\ast} ) \{ \hat{q}_\mathrm s,\Phi_\sigma \} + i\sqrt{d_k'} (\alpha_k + \alpha_k^\ast) [\hat{q}_\mathrm s, \Phi_\sigma] \\
    + \frac{d_k''}{\sqrt{d_k'}}(\partial_{\alpha_k^\ast}\Phi_\sigma\hat{q}_\mathrm s+\hat{q}_\mathrm s\partial_{\alpha_k}\Phi_\sigma)\, .
\end{multline}

By way of example, we specify this general expression for the Wigner representation ($\sigma=0$). In terms of real coordinates ($p_k,q_k$) via $\alpha_k^{(\ast)} = (q_k \pm ip_k)/\sqrt{2}$, we obtain\footnote{Considering the normalization factor $1/\sqrt{2}$, we have the relations $\partial_\alpha = (\partial_q-i\partial_p)/\sqrt{2}$, $\partial_{\alpha^\ast} = (\partial_q+i\partial_p)/\sqrt{2}$.}
\begin{eqnarray}\label{Eq:es-evolve}
\label{Eq:wigner}
    \dot{\Phi}_0 = \mathcal{L}_0\Phi_0 &+& \sum_k \gamma_k\frac{\partial}{\partial q_k} \left(q_k + \frac{1}{2}\frac{\partial}{\partial q_k} \right) \Phi_0 \nonumber\\
   & + &\sum_k \gamma_k\frac{\partial}{\partial p_k} \left(p_k + \frac{1}{2}\frac{\partial}{\partial p_k} \right)\Phi_0\;\;,
\end{eqnarray}
with  the system generator in extended space obtained as
\begin{multline}
\label{Eq:es-phase}
    \mathcal{L}_0\Phi_0 = -i[\hat{H}_\Phi, \Phi_0]+\sum_k \{h_k,\Phi_0\}_\mathrm{pq}   \\
- \sum_k \left(\frac{d_k'}{\sqrt{2 d_k'}}\frac{\partial}{\partial p_k} - \frac{d_k''}{\sqrt{2d_k''}}\frac{\partial}{\partial q_k}\right) \{\hat{q}_\mathrm{s}, \Phi_0\} \;\;.
\end{multline}
Here, the bare dynamics is generated by an effective Hamiltonian
\begin{equation}\label{Eq:es-sub}
\hat{H}_\Phi=\hat{H}_\mathrm s-\hat{q}_\mathrm s\sum_k \left(\sqrt{2 d_k'} q_k -\frac{d_k''}{\sqrt{2 d_k'}} \frac{\partial}{\partial p_k}\right) \, 
\end{equation}
for the system degree of freedom and classical Hamiltonians
\begin{equation}
h_k= \frac{\omega_k}{2} \left( p_k^2+q_k^2\right)
\end{equation}
for the reservoir modes in phase space, where, for a given Hamiltonian $h$, a phase space density $\rho$ evolves according to the classical Liouville equation $\dot{\rho}=\{h, \rho\}_\mathrm{qp}$ with Poisson brackets $\{\cdot,\cdot\}_\mathrm{qp}$ (while anti-commutators $\{\cdot,\cdot\}$ carry no index). The system generator in extended space (\ref{Eq:es-phase}) also contains coupling terms between system and reservoir modes that appear as drift terms with respect to position and momentum. 
By setting $d_k''=0$, however, this symmetry is broken. While isolated reservoir modes follow in phase space representation purely classical dynamics, the bare (undamped) time evolution of the full compound displays features which must be ascribed to quantum dissipation.

Damping dependence of the modes comes into play through the time evolution for the Wigner function in (\ref{Eq:wigner}), where it appears symmetrically in the phase space variables with identical coefficients $\gamma_k$, very different to what is known for conventional classical Fokker-Planck equations, 
where damping and diffusion manifest themselves only in terms related to momentum \cite{risken84,lucke2001dissipative,maier2010quantum,ankerhold01,ankerhold03,ankerhold2004low,skinner1979derivation,shi2009electron}. 
The structure of these operators resembles the structure known from the classical Smoluchowski equation for harmonic systems with dimensionless diffusion of 1/2 \cite{risken84}. Note though that the coefficients $\gamma_k$ as well as the frequencies $\omega_k$ and amplitudes $d_k$ depend implicitly on temperature (cf. (\ref{Eq:bcfd})). 

To arrive at the physical reduced density operator, a straightforward way is to integrate out $(q_k,p_k)$ degrees of freedom in the space spanned by the non-orthogonal basis
\begin{equation}
\begin{split}
    \chi_{\bf m,n}^{\mathcal{S}} & = \mathcal{S}^{-1}\chi_{\bf m,n}  \\ 
    &= \sum_{\bf r} \frac{{\bf m}!\, {\bf n}!}{\bm{r}! ({\bf m-r})!\, ({\bf n-r})!} \,  \chi_{\bf m-r, n-r}   \;\;
\end{split}    
\end{equation}
with Hermite polynomials $\chi_{\bf m,n}(q,p)=\chi_{\bf m}(q)\chi_{\bf n}(p)$  \cite{risken84}. Then, the reduced density matrix can be calculated as
\begin{equation}
\begin{split}
    \hat{\rho}_\mathrm s(t) &= \int \prod_k\mathrm dq_k \mathrm dp_k\ \Phi_0 (\{q_k\},\{p_k\}, t) \\
     &=\sum_{\bf m,n} \hat{\rho}_{\bf m,n}(t) \int \prod_k\mathrm dq_k \mathrm dp_k\, \chi_{\bf m,n}^{\mathcal{S}} \\
     & = \sum_{\bf n} \hat{\rho}_{\bf n,n}(t)  \;\;.
\end{split}
\end{equation}
in agreement with (\ref{Eq:recal-rho}).

The Wigner formulation also provides (as the QD-MESS) access to reservoir observables  according to 
\begin{eqnarray}
    \langle \hat{\mathcal{O}} \rangle &=& \sum_{\bf m,n} \hat{\rho}_{\bf m,n}(t) \int\prod_k \mathrm dq_k \mathrm dp_k\, \mathcal{O}(\{q_k\},\{p_k\}) \nonumber\\
    &&\hspace{3cm}\times \chi_{\bf m,n}^{\mathcal{S}} (\{q_k\},\{p_k\}) \;\; 
\end{eqnarray}
with the Wigner representation of the reservoir operator $\hat{\mathcal{O}}$. 

At this point, it is again interesting to bridge these exact  results to approximate formulations which have been developed previously, see Fig.~\ref{fig1}. 
For example, by setting $\omega_k = 0$ in (\ref{Eq:es-phase}), one assumes that correlation functions decay purely exponential in time, an approximation that is justified as long as bath memory times follow the classical Onsager regression theorem, i.e., for quantum dynamics that appears, at least on a coarse grained time scale, as classical-like \cite{risken84,lucke2001dissipative,maier2010quantum,ankerhold01,ankerhold03,ankerhold2004low,skinner1979derivation}. In the opposite situation, where $\gamma_k=0$ and additionally $d_k \in \mathbb{R}^+$, one arrives at the interesting situation that Eqs. (\ref{Eq:es-evolve} - \ref{Eq:es-sub}) describe the bare dynamics in an extended quantum-classical state space. This way, one regains the mixed quantum-classical Liouville (MQCL) equation is obtained, commonly applied in chemical physics \cite{shi04b,liu14,yan2020new,kapral99,kapral01,kapral06,hsieh2012nonadiabatic}.

\section{Relation to stochastic schemes}
\label{sec:section5}

In the past, non-perturbative approaches for quantum dissipation have been formulated based on unraveling schemes introducing stochastic auxiliary fields with properties determined by the reservoir correlation $C(t)$. The best-known are the Stochastic Liouville-von Neumann Equation (SLN) and the non-Markovian Stochastic Schrödinger Equation (nonM-SSE). While their original derivations differ, they both operate with wave functions and density matrices which are time evolved in presence of a specific noise realization.  The physical density matrix appears as an average over a sufficiently large ensemble. Recently, to boost the performance, the nonM-SSE has been combined with hierarchy schemes to the Hierarchy of Stochastic Pure States (HOPS). We here demonstrate that the framework formulated in Sec.~\ref{sec:Liouville-Fock} also provides a uniform platform to derive these latter approaches (cf. ~Fig.~\ref{fig1}). 

\subsection{Stochastic unraveling: Stochastic Liouville-\\von Neumann Equation}
\label{sec:sln}
To start, we first introduce a time-dependent similarity transformation in Fock-space according to
\begin{equation}
    \mathcal{S} = \prod_{k}\exp\{-\Gamma_k t + \mu_k\hat{m}_k + \nu_k\hat{n}_k\} \;\;,
\end{equation}
where parameters $\mu_k$, $\nu_k$ are determined via  $e^{\mu_k} = i\sqrt{2}$ and $e^{\nu_k} = -i\sqrt{2}$, respectively. The action of $\mathcal{S}$ on the creation-annihilation operators simply reads 
\begin{subequations}\label{Eq:time-similarity}
\begin{align}
\mathcal{S}\hat{a}_k^{(\dagger)}\mathcal{S}^{-1} = {\rm e}^{\mp(\mu_k + z_k t )}\hat{a}_k^{(\dagger)}\,\, \\
\mathcal{S}\hat{b}_k^{(\dagger)}\mathcal{S}^{-1} = {\rm e}^{\mp(\nu_k+z_k^\ast t)}\hat{b}_k^{(\dagger)}\,.
\end{align}    
\end{subequations}
This time-dependent transformation maps onto a rotating frame in Fock-space in which time-dependent operators are defined as  $\hat{a}_k^{(\dagger)}(t)= {\rm e}^{\mp z_k t}\hat{a}_k^{(\dagger)}$ and $\hat{b}_k^{(\dagger)}(t)= {\rm e}^{\mp z_k^\ast t}\hat{b}_k^{(\dagger)}$.
Now, upon implementing the above transformation (\ref{Eq:time-similarity}) on Eq. (\ref{Eq:standard-heom-liouville-fock}), one arrives with $\hat{\rho}_{\mu} = \mathcal{S}\hat{W}$ in Liouville space at
\begin{equation} \label{Eq:sln1}
    \dot{\hat{\rho}}_{\mu} = -i\mathcal{L}_\mathrm{s} \hat{\rho}_{\mu} + i\left( [\hat{q}_s, \hat{\xi}(t) \hat{\rho}_{\mu}] + \{\hat{q}_s, \frac{\hat{\nu}(t)}{2}\hat{\rho}_{\mu}\} \right)\, ,
\end{equation}
where collective reservoir operators
\begin{subequations}\label{Eq:combined-noise}
 \begin{equation}
     \hat{\xi}(t) = \sum_k\{\frac{\sqrt{d_k}}{2}(\hat{a}_k(t) +\hat{a}_k^\dagger(t)) +  \frac{\sqrt{d_k^\ast}}{2}(\hat{b}_k(t) + \hat{b}_k^\dagger(t)) \} 
 \end{equation}
 \vspace*{-0.5cm}
 \begin{equation}
     \hat{\nu}(t) = \sum_k\sqrt{2d_k}\hat{a}_k^\dagger(t) + \sqrt{2d_k^\ast}\hat{b}_k^\dagger(t) \;\;
 \end{equation}
\end{subequations}
capture the complete influence of the reservoir modes. Their properties follow by taking expectation values with respect to the initial state of the reservoir as 
$\langle \hat{\xi}(t) \hat{\xi}(\tau)\rangle = \mathrm{Re}\,C(t-\tau)$, $\langle \hat{\xi}(t) \hat{\nu}(\tau)\rangle = 2i\,\Theta(t-\tau)\, \mathrm{Im}\,C(t-\tau)$ with $\Theta(\cdot)$ being a step function, and $\langle \hat{\nu}(t)\hat{\nu}(\tau)\rangle = 0$. Since we are only interested in the effective impact of the Gaussian reservoir modes on the system of interest, these correlations may as well be constructed by introducing classical fluctuating fields $\mu = \{\xi,\nu \}$ that obey the same statistics as their quantum counterparts. Substituting the respective operator combinations in (\ref{Eq:combined-noise}) with these classical fields yields an effective time evolution equation known as the Stochastic Liouville-Von Neumann (SLN) equation for a density operator $\rho_{\xi, \nu}$ \cite{stockburger02}, i.e.
\begin{multline}
    \frac{d}{dt} \hat{\rho}_{\xi, \nu} = -i\mathcal{L}_s\hat{\rho}_{\xi, \nu}
 +   i \xi(t) [\hat{q}_\mathrm s, \hat{\rho}_{\xi, \nu}] + i \frac{\nu(t)}{2} \{ \hat{q}_\mathrm s, \hat{\rho}_{\xi, \nu} \} \;\; .
\end{multline}
This SLN provides the same reduced density operator as the one determined by Eq.~(\ref{Eq:sln1}) when mean values with respect to the corresponding classical probability distribution are taken, i.e.\  $\hat{\rho}_s(t)=\langle \mathbf{0, 0}| \mathcal{S}^{-1}\hat{\rho}_\mu(t)\rangle=\mathbf{M}[\hat{\rho}_{\xi, \nu}]$.
In essence, the SLN thus emerges, up to a scaling, as the representation of the QD-MESS (\ref{Eq:standard-heom-liouville-fock}) in a rotating Fock-frame and with combinations of mode operators replaced by classical noise fields.

\subsection{Stochastic Unraveling: Hierarchy of Pure States}
\label{subsec:hops}
To unravel the QD-MESS given by (\ref{Eq:standard-heom-density}) into a Schr\"odinger wave equation, we start by re-expressing operator terms in the coherent state representation, namely,
\begin{subequations}
\begin{align}
    \sqrt{d_k}\hat{a}_k\hat{\rho}\hat{q}_\mathrm s &\leftrightarrow \sqrt{d_k}\phi_k\hat{\rho}\hat{q}_\mathrm s\\
    \sqrt{d_k^\ast}\hat{q}_\mathrm s \hat{\rho} \hat{a}_k^\dagger &\leftrightarrow \sqrt{d_k^\ast}\hat{q}_\mathrm s \psi_k \hat{\rho}
\end{align}
\end{subequations}
with the correlation function satisfying the relations
\begin{subequations}
\begin{align}
\sum_{k,j} \sqrt{d_k} \sqrt{d_j}\langle \phi_k^\ast(t)\phi_j(\tau) \rangle &= C(t-\tau)\,\Theta(t-\tau) \\
\sum_{k,j} \sqrt{d_k^\ast} \sqrt{d_j^\ast}\langle \psi_k^\ast(t)\psi_j(\tau) \rangle &= C^\ast(t-\tau)\,\Theta(t-\tau) \;\;.
\end{align}
\end{subequations}

Now, the above correlation function $C^\ast(t-\tau)$ can also be constructed by introducing complex-valued fields $Z_t^\ast = \sum_k\sqrt{d_k^\ast} \psi_k^\ast(t)$ and  $Z_t = \sum_k\sqrt{d_k^\ast} \psi_k(t)$ \footnote{The scaling $\sqrt{d_k^\ast}$ origins from the way to assign the amplitude $d_k^\ast$ as in (\ref{Eq:phik}); here the fields $Z_t^\ast$ and $Z_t$ appear thus not as conjugate pairs.}. Thus, the QD-MESS (\ref{Eq:standard-heom-density}) can be reformulated as the following stochastic Schr\"odinger equation
\begin{multline}
\partial_t |\Psi_Z\rangle = -i\hat{H}_\mathrm s |\Psi_Z\rangle + \hat{q}_\mathrm s Z_t |\Psi_Z\rangle \\
-\sum_k [z_k\hat{a}_k^{\dagger}\hat{a}_k + \sqrt{d_k}\hat{q}_\mathrm s (\hat{a}_k - \hat{a}_k^\dagger)] |\Psi_Z\rangle \;\;,
\end{multline}
with an unraveled density matrix $\rho=|\Psi_Z\rangle\langle\Psi_Z|$, where $|\Psi_Z\rangle$ denotes a state in the mixed system Hilbert-Fock state space depending on a stochastic field $Z_t$. The statistics of this field obeys $\langle Z_t^\ast Z_{\tau}\rangle = C^\ast(t-\tau)\Theta(t-\tau)$ and $\langle Z_t Z_\tau\rangle = \langle Z_t^\ast Z_\tau^\ast\rangle = 0$.

By projecting onto the Fock space such that $|\Psi_Z(t)\rangle = \sum_k \psi_{\bf n}[Z_t] |{\bf n}\rangle$, we recover, up to a scaling, the Hierarchy of Pure States (HOPS) representation \cite{suess14} 
\begin{multline}
    \partial_t\psi_{\bf n} = -i\hat{H}_s\psi_{\bf n} +\hat{q}_\mathrm s Z_t \psi_{\bf n} -\sum_{k=1}^K\left[ n_kz_k \psi_{\bf n}  \right.\\ \left. -\sqrt{n_kd_k}\hat{q}_\mathrm s\psi_{\bm{n}_k^-} + \sqrt{(n_k +1)d_k}\hat{q}_\mathrm s\psi_{\bm{n}_k^+} \right] \;\;.
\end{multline}
Likewise, one can obtain a corresponding HOPS representation by unraveling the Lindblad-type equivalent (\ref{Eq:standard-heom-lindblad}).

\section{Overcoming Computational Challenges: Leveraging Matrix Product States}
\label{sec:section6}

As is well-known in all treatments of open quantum dynamics, simulations become increasingly demanding computationally for very low temperatures, higher dimensionality of system Hilbert spaces, and structured reservoir spectral densities. As we are argued in the last sections, the QD-MESS provides a very efficient toolbox to tackle these challenges. Within the pool of conventional HEOM approaches, previously progress has already been achieved with the Ishizaki-Tanimura truncation \cite{ishizaki05} and on-the-fly filtering \cite{shi2009efficient}. The QD-MESS (\ref{Eq:standard-heom-liouville-fock}) equation, or its representation in form of the FP-HEOM, as detailed in (\ref{Eq:standard-heom}), allows for another significant boost by employing tensor network states \cite{shi2018efficient,yan2021efficient,borrelli2019density,ke2023tree}. This enhancement is especially advantageous not only when the Hilbert space of auxiliary bosonic modes grows but also when the dimensionality of the system Hilbert space expands. Here, we give a concise account of the implementation. 

By transforming (\ref{Eq:standard-heom-liouville-fock}) into a tensor network representation using MPS \cite{borrelli2019density,yan2021efficient}, computational resources can be scaled linearly with an increasing number of auxiliary bosonic modes. This is feasible as ADOs in open quantum systems exhibit low entanglement during time evolution. Bond dimensions can be systematically incremented to verify convergence.

Expressing Eq. (\ref{Eq:standard-heom}) as $\partial_t\hat{\rho}_{\bf m,n}(t) = -i\mathcal{L}_{\rm FP} \hat{\rho}_{\bf m,n}(t)$, we propose that the multidimensional array $\hat{\rho}_{\bf m,n}$ can be recast into an open boundary conditions matrix product state \cite{shi2018efficient} (also known as a tensor train \cite{lubich15}), as shown below:
\begin{equation}
\begin{split}
\rho_{\mathbf{m,n}}^{i j} 
\approx & B_0(i)B_1(m_1)B_2(n_1) \\
        & \cdots B_{2K-1}(m_K) B_{2K}(n_K) B_{2K+1}(j) \;\;,
\end{split}
\end{equation}
where the superscripts $i,j$ are associated with the degrees of the subsystem. The MPS cores $B_k$ are three dimensional arrays with ranks $r_{k-1}\times s_{k-1}\times r_k$ and $r_{-1} = r_{2K+2} = 1$ with $r_k$ being the bond dimension and $s_k$ being the maximal occupation on a local site.

In analogy to the MPS, every operator can be expressed as a matrix-product operator (MPO), namely as a contraction of rank-4 tensors $\mathcal{M}_k$, i.e.,
\begin{equation}
\begin{split}
   -i\mathcal{L}_{\rm FP}  
   =& \mathcal{S}_L\otimes \mathcal{M}_1\otimes \mathcal{M}_2\ldots\mathcal{M}_{2k-1}\otimes\mathcal{M}_{2k} \\
    & \ldots\mathcal{M}_{2K-1}\otimes\mathcal{M}_{2K}\otimes\mathcal{S}_R \;\;.
\end{split}
\end{equation}
with  $\mathcal{M}_{2k-1}$ and $\mathcal{M}_{2k}$ being associated with sites $m_k$ and $n_k$ as
\begin{equation}
\mathcal{M}_{2k-1} = 
\begin{bmatrix}
       \hat{I}_{k} & 0 & 0 & 0\\
-z_{k}\hat{a}_{k}^{\dagger}\hat{a}_{k} & \hat{I}_{k} & \sqrt{d_k}\hat{a}_k & 0\\
 0 & 0 & \hat{I}_{k} & 0 \\
\sqrt{d_k} (\hat{a}_k^{\dagger}-\hat{a}_k)  & 0 & 0 & \hat{I}_k
    \end{bmatrix} \;\;;
\end{equation}
 \begin{equation}
\mathcal{M}_{2k} = 
\begin{bmatrix}
       \hat{I}_{k} & 0 & 0 & 0\\
-z_k^*\hat{b}_k^{\dagger}\hat{b}_k & \hat{I}_{k} & \sqrt{d_k^*}(\hat{b}_{k}^{\dagger} - \hat{b}_k) & 0 \\
       0 & 0 & \hat{I}_{k} & 0 \\ \sqrt{d_k^*}\hat{b}_k  & 0 & 0 &\hat{I}_k
    \end{bmatrix} \;\;.
\end{equation}

The tensors $\mathcal{S}_{L/R}$ correspond to subsystem through
\begin{equation}
\begin{split}
    \mathcal{S}_L &=  \begin{bmatrix}
       -i\hat{H}_\mathrm s & \hat{I}_\mathrm s & 0 & \hat{q}_\mathrm s
    \end{bmatrix} \;\;; \\
    \mathcal{S}_R^{\rm T} &=  \begin{bmatrix}
        \hat{I}_\mathrm s & i\hat{H}_\mathrm s & \hat{q}_\mathrm s & 0
    \end{bmatrix}  \;\;.
\end{split}
\end{equation}

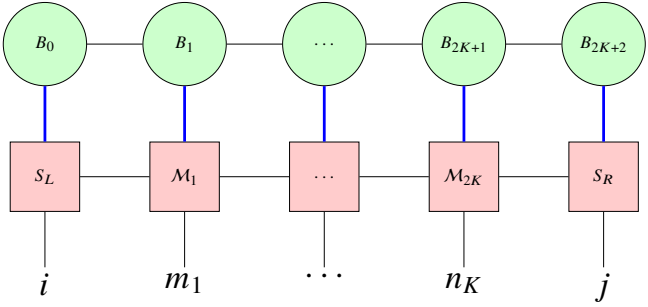
\begin{figure}[!h]
\centering
\resizebox{0.48\textwidth}{!}{\begin{tikzpicture}[scale=1,
  block/.style={draw, fill=green!20, minimum size=1.5cm, circle},
  tensor/.style={draw, fill=red!20, minimum size=1.25cm, rectangle},
  plain node/.style={font=\LARGE}
]

\node[block] (B0) at (0,3) {$B_0$};
\node[block, right=of B0] (B1) {$B_1$};
\node[block, right=of B1] (Dots) {$\ldots$};
\node[block, right=of Dots] (B2K1) {$B_{2K+1}$};
\node[block, right=of B2K1] (B2K2) {$B_{2K+2}$};

\node[tensor, below=of B0] (SL) {$S_L$};
\node[tensor, below=of B1] (M1) {$\mathcal{M}_1$};
\node[tensor, below=of Dots] (DotsR) {$\ldots$};
\node[tensor, below=of B2K1] (M2K) {$\mathcal{M}_{2K}$};
\node[tensor, below=of B2K2] (SR) {$S_R$};

\node[plain node, below=of SL] (i) {$i$};
\node[plain node, below=of M1] (m1) {$m_1$};
\node[plain node, below=of DotsR] (DotsP) {$\ldots$};
\node[plain node, below=of M2K] (nK) {$n_K$};
\node[plain node, below=of SR] (j) {$j$};

\draw (B0)--(B1)--(Dots)--(B2K1)--(B2K2);
\draw (SL)--(M1)--(DotsR)--(M2K)--(SR);

\foreach \c/\r in {B0/SL, B1/M1, Dots/DotsR, B2K1/M2K, B2K2/SR} {
    \draw[blue, line width=0.5mm] (\c)--(\r);
}

\foreach \r/\p in {SL/i, M1/m1, DotsR/DotsP, M2K/nK, SR/j} {
    \draw (\r)--(\p);
}
\end{tikzpicture}}
\vspace*{-0.5cm}
\caption{ Graphical notation of QD-MESS (\ref{Eq:standard-heom-liouville-fock}). The upper connected circles: MPS representation of ADOs; while the lower connected triangles: MPO representation of the total operator. The MPO acts on the MPS via tensor contraction denoted by blue lines.}
\label{fig3}
\end{figure}

It can be demonstrated that $[\mathcal{M}_j,\mathcal{M}_k] = 0$, signifying that all auxiliary modes possess topological equivalence. A visual representation of the MPS for (\ref{Eq:standard-heom-liouville-fock}) is illustrated in Fig. \ref{fig3}.

To conclude this part, we offer a technical remark. In equation (\ref{Eq:standard-heom-liouville-fock}), the Liouvillian $\mathcal{L}_{\rm FP}$ couples the system's degrees of freedom to a collection of non-interacting auxiliary modes, corresponding to a "star" topology. This "star"-like Liouvillian induces long-range interactions between the system and the auxiliary modes, rendering conventional DMRG methods ineffective \cite{daley2004time,schollwock2005the,schollwock11}. However, based on the time-dependent MPO-MPS representation, an efficient propagation scheme can be implemented through the time-dependent variational principle (TDVP) algorithm \cite{haegeman2011time,lubich15,haegeman2016unifying,yang2020time}.

\section{Perturbative treatments}
\label{sec:perturbative}
The QD-MESS formulation presented so far offers also ways for approaches which in some or the other way rely on perturbative treatments. While they are thus limited in their applicability, they nevertheless are often elegant and powerful alternatives to full simulations. Here, we discuss first how the conventional HEOM is regained due to a simplification of the decomposition (\ref{Eq:bcfd}) of the reservoir correlation, and then turn to a systematic weak coupling expansion, see Fig.~\ref{fig1}. 

\subsection{Reduction to conventional HEOM}

By setting $\omega_k = 0$ in Eq. (\ref{Eq:bcfd}) so that $z_k = z_k^\ast = \gamma_k$, all propagators $\mathcal{G}_k$ and its conjugation $\mathcal{G}^\ast_k$ are degenerate. Thus,  the influence functional (\ref{Eq:ifk}) simplifies to read
$F[q_\mathrm s^+,q_\mathrm s^-]= \sum_{k=1}^K \tilde{F}_k[q_\mathrm s^+,q_\mathrm s^-]$ with 
\begin{equation}
\tilde{F}_k[q_\mathrm s^+,q_\mathrm s^-] = \langle q_\mathrm s^+-q_\mathrm s^-,-\mathcal{G}_k[d_kq_\mathrm s^+ - d_k^\ast q_\mathrm s^-] \rangle\, .
\end{equation}
Upon introducing auxiliary coherent state pairs $(\phi_k^\ast,\phi_k)$ according to $\hat{a}_k|\phi_k\rangle = \phi_k |\phi_k\rangle$, one arrives at
\begin{equation}
e^{\tilde{F}_k[q_\mathrm s^+, q_\mathrm s^-]} = \int \mathcal D[\phi_k^\ast, \phi_k] \, \mathrm{e}^{i \tilde{S}_k[\phi_k^\ast,\phi_k,q_s^+,q_s^-]} \;\;
\end{equation}
with time local action
\begin{equation}
    \tilde{S}_k[\phi_k^\ast,\phi_k,q_s^+,q_s^-] = \int_0^t d\tau\, [i\phi_k^\ast\partial_\tau \phi_k - \tilde{H}_k(\phi_k^\ast,\phi_k)] \;\;
\end{equation}
and 
\begin{equation}
    \tilde{H}_k = i\gamma_k\phi_k^\ast\phi_k - i(q_\mathrm{s}^+ - q_\mathrm{s}^-) \phi_k + i (d_k q_\mathrm{s}^+ - d_k^\ast q_\mathrm{s}) \phi_k^\ast \;\;.
\end{equation}

Now, following similar steps as discussed above, yields the conventional HEOM as originally derived by Tanimura and Kubo \cite{tanimura89,tanimura06,tanimura2020numerically}
\begin{multline}\label{Eq:taka-heom}
    \dot{\hat{\rho}}_{\bf n} = -i\mathcal{L}_\mathrm s\hat{\rho}_{\bf n} + \sum_k \left[ -n_k\gamma_k\hat{\rho}_{\bf n} + \sqrt{n_k + 1}\,[\hat{q}_\mathrm s, \hat{\rho}_{\bm{n}_k^+}] \right.\\ \left.
    - \sqrt{n_k}\, (d_k\hat{q}_\mathrm s\hat{\rho}_{\bm{n}_k^-} - d_k^\ast\hat{\rho}_{\bm{n}_k^-}\hat{q}_\mathrm s ) \right] \;\;.
\end{multline}

However, in the conventional HEOM this simplification comes with a severe drawback: 
Namely, in absence of a suitable rational approximation one sets $\gamma_k=\nu_k$ with the Matsubara frequencies $\nu_k=2\pi k/\beta$. This implies that at lowering temperatures the number of terms $K$ [Eq.  (\ref{Eq:bcfd})] increases dramatically and grows without bound for $T\to 0$ \cite{mark1}. Accordingly, the number of ADOs explodes, roughly according to  $\exp[L\, {\rm ln}(K/L)]/\sqrt{L}$ for $K\gg L$ \cite{shi2009efficient}, where $L= \sum_{k=1}^K n_k$ denotes the hierarchy level of truncation. In general, the conventional HEOM is thus limited to sufficiently elevated temperatures and smooth reservoirs. One can show though that in specific cases, e.g.\ for sufficiently smooth spectral bath densities, this problem can be cured by using a proper decomposition of $C(t)$ based on a rational approximation with real-valued $z_k$. 

\subsection{Systematic weak coupling expansion}
\label{subsec:weak-coupling}
The best-known expansion of the formal expression of the reduced density operator in terms of the system-reservoir coupling is the set of approximations that lead to Born-Markov Master and Lindblad equations. In this section, we describe a systematic procedure for a weak coupling perturbation series in the framework of the extended state space introduced in Sec.~\ref{sec:unraveling} which does avoid the Markov approximation. 

We start with the observation that the time evolution equation for $\hat{W}$ in extended Liouville-Fock space (\ref{Eq:standard-heom-liouville-fock}) suggests to define effective bi-linear interactions, i.e., 
\begin{eqnarray}
\hat{V}_k \hat{W} &=& \sqrt{d_k} \hat{q}_\mathrm s a_k^\dagger \hat{W} - \sqrt{d_k} [\hat{q}_\mathrm s, a_k\hat{W}] \nonumber \\
\hat{\bar{V}}_k \hat{W} &=& \sqrt{d_k^\ast} b_k^\dagger \hat{W} \hat{q}_\mathrm s + \sqrt{d_k^\ast} [\hat{q}_\mathrm s, b_k \hat{W}] \ .
\end{eqnarray}
The density in the extended Liouville-Fock space $\hat{W}$ [see Eqs. (\ref{Eq:fullaction}) and (\ref{Eq:standard-heom-liouville-fock})] can then be Taylor expanded with respect to $\hat{V} = \sum_k (\hat{V}_k + \hat{\bar{V}}_k)$. This expansion takes the elegant form of a Dyson relation in functional space, namely, 
\begin{equation}\label{Eq:dyson}
    \hat{W}(t) = \hat{G}_0(t,0) \hat{W}(0) +\int_0^t d\tau\, \hat{G}_0(t,\tau) \hat{V}(\tau) \hat{W}(\tau)\;
\end{equation}
Here, $\hat{G}_0(t,\tau)$ denotes the bare system-bath propagator in extended Liouville-Fock space (i.e.\ setting the system-bath coupling $\hat{H}_{\rm sb}=0$).

In extended space the retardation of reservoir modes is encoded in the 'damping' rates ${\rm Re}(z_k)\equiv \gamma_k$:  The memory with respect to sluggish modes (small $\omega_k$ and $\gamma_k$ compared to typical system frequencies) is long-ranged and thus gives rise to strongly non-Markovian behavior. In contrast, the fast modes (large $\omega_k$ and $\gamma_k$) can be safely truncated or even treated approximately Markovian on a coarse grained time scale. Here, we do not follow this idea, but proceed with writing down an expansion to second order in $\hat{V}$ in Eq. (\ref{Eq:dyson})  followed by tracing out FP-modes then leads to a generalized Redfield master equation \cite{yan2021theoretical,xu2021heat,xu2022minimal}, in the interaction picture given by
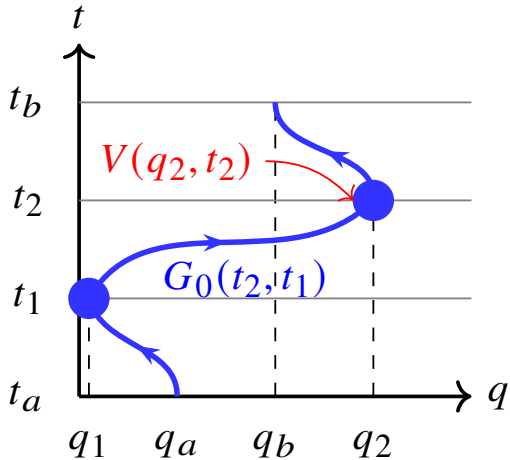
\begin{figure}[!h]
\centering
\resizebox{0.4\textwidth}{!}{\begin{tikzpicture}[scale=1.5]

\draw[->,thick] (0,0) -- (2,0) node[right] {$q$};
\draw[->,thick] (0,0) -- (0,1.8) node[above] {$t$};

\draw[black, dashed, line width =0.4pt] (0.05,0) -- (0.05,0.5);
\draw[black, dashed, line width =0.4pt] (1,0) -- (1,1.5);
\draw[black, dashed, line width =0.4pt] (1.5,0) -- (1.5,1);

\draw[solid,black!50] (0,0.5) -- (2,0.5);
\draw[solid,black!50] (0,1) -- (2,1);
\draw[solid,black!50] (0,1.5) -- (2,1.5);

\draw (1/2,-0.1) node[below] {$q_a$} (1,-0.1) node[below] {$q_b$} (0.05,-0.1) node[below] {$q_1$} (1.5,-0.1) node[below] {$q_2$};
\draw (-0.1,0.5) node[left] {$t_1$} (-0.1,1) node[left] {$t_2$} (-0.1,1.5) node[left] {$t_b$} (-0.1, 0) node [left] {$t_a$};

\fill[blue!80] (0.05,0.5) circle (3pt) (1.5,1) circle (3pt);

\node[red] at (0.5,1.2) {$V(q_2,t_2)$};
\draw[red, ->] (0.95, 1.2) to [out=0, in=135] (1.4,1);
\node[blue] at (0.85,0.6) {$G_0(t_2,t_1)$};

\draw[blue!80, line width=1.2pt, postaction={decorate, decoration={markings, mark=at position 0.5 with {\arrow[blue!80]{Stealth[length=5pt]}}}}] (0.5,0) to[out=90,in=-75] (0.05,0.5);
\draw[blue!80, line width=1.2pt, postaction={decorate, decoration={markings, mark=at position 0.5 with {\arrow[blue!80]{Stealth[length=5pt]}}}}] (0.05,0.5) to[out=60,in=-135] (1.5,1);
\draw[blue!80, line width=1.2pt, postaction={decorate, decoration={markings, mark=at position 0.5 with {\arrow[blue!80]{Stealth[length=5pt]}}}}] (1.5,1) to[out=85,in=-90] (1,1.5);

\end{tikzpicture}}
\vspace*{-0.5cm}
\caption{Diagram of the propagation of the extended state $W(t)$ according to the Dyson equation (\ref{Eq:dyson}) when represented in position (system) and coherent states (effective reservoir modes) with\ $q = \{q_\mathrm s,\phi_k,\phi_k^\ast,\psi_k,\psi_k^\ast\}$. Blue dots indicate the interaction potential $V$, while the blue paths represent the bare propagator of the compound.}
\label{fig4}
\end{figure}
 
\begin{multline}\label{redfieldplus}
    \dot{\hat{\rho}}_\mathrm s(t) = -\int_0^t d\tau\, \left\{ [\hat{q}_\mathrm s(t),C(t-\tau)\hat{q}_\mathrm s(\tau)\hat{\rho}_\mathrm s(\tau)] \right. \\ \left.
    - [\hat{q}_\mathrm s(t), C^\ast(t-\tau)\hat{\rho}_\mathrm s(\tau)\hat{q}_\mathrm s(\tau)] \right\}\;\;.
\end{multline}
 The most appealing point is that within the FP-HEOM this equation appears quite naturally when truncating the full hierarchy of nested equations of motion {\em after tier 1}, i.e.\ to set all ADOs with indices $\sum_k (n_k+ m_k) > 1$ to zero. In contrast to the standard Redfield equation, here, the density operator on the right hand side is taken at all intermediate times. We thus assign to the generalized Redfield equation (\ref{redfieldplus}) the name Redfield+.  The conventional Redfiled is regained on a coarse-grained time scale for sufficiently slow relaxation via $\rho_\mathrm s(\tau)\to \rho_\mathrm s(t)$.

In principle, higher order terms based on the formulae in (\ref{Eq:dyson})  can easily be derived formally, however, practical calculations are cumbersome as they involve  large numbers of multidimensional integrals that are very sensitive to numerical errors. In fact, it is the elegance of the HEOM formulation that it turns time-ordered integrals into a nested set of ordinary differential equations. 

\section{Conclusion and Remarks}
\label{sec:section8}

The Feynman-Vernon path integral offers an exact framework to capture the temporal propagation of the reduced density matrix. However, its practical implementation for direct computational approaches encounters severe challenges due to the dynamical sign problem, and the time non-locality renders the direct conversion into tractable equations of motion impossible. Hence, various hybrid methods partially 'unraveling' the non-locality in time have been developed recently aiming at balancing numerical stability and accuracy through time-local evolution equations.

The present work demonstrates that these methods evolve from a common platform formulated in Liouville-Fock space and characterized by a minimal set of auxiliary bosonic modes (QD-MESS). These modes naturally appear when the central ingredient for the path integral expression, the spectral noise power, is systematically represented in complex frequency space via barycentric rational functions. In the time domain this leads to a decomposition of the reservoir correlation in terms of harmonic modes with complex valued frequencies and amplitudes which in turn allows for an unraveling of the influence functional in terms of coherent states. As a consequence, the exact open system dynamics can be mapped onto a time-local dynamics in a minimally extended Liouville-Fock space. This QD-MESS serves as a universal framework to derive alternative representations via similarity transformations and new and known approximate treatments via restrictions on mode amplitude and/or frequencies.
This not only provides a solid foundation of their derivation but also emphasizes the versatility of the QD-MESS. 

The FP-HEOM appearing as the QD-MESS in the occupation number representation can be efficiently propagated with Matrix Product States (MPS). Thereby it merges computational efficiency with longtime scales stability, high precision, and wide-range applicability across all temperatures and arbitrary reservoir spectral noise power. The decomposition of the reservoir auto-correlation function is system-agnostic and solely demands the noise power, obtained from experimental data, as input.
The QD-MESS also allows for the calculation of mixed expectation values (system+reservoir) as well as certain reservoir observables, offering insight into the spreading/shrinking of entanglement in the system-reservoir compound. Practically, the QD-MESS constitutes a versatile platform to deliver highly accurate predictions for dynamical characteristic of real-world quantum technological devices.

\section{Acknowledgements}
M. X. acknowledges fruitful discussions with Qiang Shi. This work has been supported by the German Science Foundation (DFG) under AN336/12-1 (For2724), the State of Baden-W{\"u}ttemberg under KQCBW/SiQuRe, and the BMBF within the project QSolid as well as through the Academy of Finland Centre of Excellence program (project no.~336810) and THEPOW (project no.~349594), and the European Research Council under Advanced Grant no.~101053801 (ConceptQ).

\appendix
\section{Coherent-state path integral and boundary conditions}
\label{sec:CSappendix}

A detailed and logically complete characterization of the arguments put forward in Sec. \ref{sec:unraveling}, in particular, the Gaussian identity (\ref{Eq:gpi2}), must rely on a time-discrete version of the path integral, followed by the limit $h\to 0$ for the time step $h$. As a discrete representation of the scalar product (\ref{Eq:ip}), we substitute
\begin{equation}
\langle u,v\rangle \to \langle u,v \rangle_h = h \sum_{j=1}^{M-1} u^*_j v_j 
\end{equation}
where $u_j$ and $v_j$ sample $u(\tau)$ and $v(\tau)$ at times $\tau_j = j h$, $\tau_M = t$. Omission of the first and last point is both formally exact for the end result in the limit $h\to 0$ and in keeping with the fact that initial and final values of paths in a functional integral are normally considered fixed.

Similarly, we replace the integral operator $\mathcal{G}_k$ given in Eq. (\ref{Eq:convolution}) by its discrete version
\begin{equation}
\mathbf{G}_k:~v_j \mapsto h \sum_{l=1}^j  e^{-z_kh(j-l)} v_l
\end{equation}
where it is again understood that $1\leq j,l < M$. The linear map is associated with the matrix
\begin{equation}
  G_{jl}^{(k)} = \left\{
  \begin{array}{ll}
    e^{-z_kh(j-l)} & l\leq j\\
    0 & l>j
  \end{array}
  \right. .
\end{equation}
This upper diagonal matrix is obviously regular, and its inverse is
\begin{align}
  \Delta_{jl}^{(k)} &= \left\{ \begin{array}{ll}
  h^{-1} \delta_{jl} & j=1\\
  h^{-1} \left(
  \delta_{jl} - e^{-z_kh}\delta_{j-1,l}\right) & j>1
  \end{array} \right. .
\end{align}
Alternatively, one can use only the upper expression with the understanding that the formally appearing zero index (for j=1) refers to a vector element with value zero.

The corresponding map can be cast into the form
\begin{equation}\label{Eq:discreteDelta}
\mathbf{\Delta}_k:~v_j \to \frac{1}{h} (v_j-v_{j-1}) + z v_{j-1} + O(h)
\end{equation}
which is a discrete version of the operator $D_k$, now with fixed boundary condition $v_0=0$.

The discrete version of the Gaussian identity (\ref{Eq:gpi2}) thus reads
\begin{multline}\label{Eq:gpi2discrete}
 \frac{1}{N}\int d^{M-1} \phi_j^{(k)} \int d^{M-1} \phi_j^{(k)}
    e^{-\langle\phi_k,\mathbf{\Delta}_k\phi_k\rangle_h + \langle w,\phi_k\rangle_h + \langle\phi_k,w'\rangle_h }\\
= e^{\langle w,\mathbf{G}_k w' \rangle_h}
\end{multline}
with
\begin{equation}
N = \int d^{M-1} \phi_j^{(k)} \int d^{M-1} \phi_j^{(k)}
  e^{-\langle\phi_k,\mathbf{\Delta}_k\phi_k\rangle_h }
\end{equation}
We can now compare Eq. (\ref{Eq:gpi2discrete}) with the coherent-state path integral~\cite{negel88} for a harmonic mode with non-hermitian Hamiltonian
\begin{equation}
    \hat{H}_k(\tau) = -i z_k a_k^\dagger a_k + i w(\tau) a_k + i w'(\tau) a_k^\dagger\;,
\end{equation}
initial state $|\phi_0^{(k)}\rangle$ and final state $|\phi_M^{(k)}\rangle$,
\begin{multline}\label{Eq:CohStateProp}
    \langle \phi_M^{(k)} | U(t) | \phi_0^{(k)}\rangle
    = \frac{1}{N} \int d^{M-1} \phi_j^{(k)} \int d^{M-1} \phi_j^{(k)} \\
    \times \exp\left(\phi_M^{(k)*}\phi_{M-1}^{(k)}-ih H(\phi_M^{(k)*},\phi_{M-1}^{(k)}\right)\\
    \times\exp\left(\sum_{j=1}^{M-1} \phi^{(k)*}_j (\phi_j^{(k)}-\phi_{j-1}^{(k)})
    -i H_k(\phi^{(k)*}_j \phi_{j-1}^{(k)}) \right),
\end{multline}
where $H_k(\psi^*,\phi) = -i z_k \psi^*\phi + i w \phi + i w' \psi^*$.
In this equation, it is now understood that the zero element in Eq. (\ref{Eq:discreteDelta}) refers to the given initial state label. Comparing the auxiliary path integral (\ref{Eq:gpi2discrete}) and the coherent-state path integral (\ref{Eq:CohStateProp}), it becomes clear that the two can be identified for $h\to 0$ if and only if
both $\phi_0^{(k)}=0$ and $\phi_M^{(k)}=0$. This leads to the conclusion that $|0\rangle$ must be both the initial and final state of the auxiliary bosons if the Feynman-Vernon action is to be reproduced.

\section{Alternative Lindblad stucture}
To complement the analysis presented in Sec.~\ref{sec:section4}, we here provide a transformation to an alternative Lindblad structure which, however, does not provide the original commutator algebra.

For this purpose, we write $ d_k = R_k\,\mathrm{e}^{i\theta_k} $ with $\theta_k$ denoting the initial phase information of the auxiliary modes.
Then, in (\ref{Eq:standard-heom-liouville-fock}) the following {\em affine} transformation is introduced  
\begin{equation}
 \begin{bmatrix}
    \hat{a}_k^\dagger \\ \hat{b}_k^\dagger \\ \hat{a}_k \\ \hat{b}_k
 \end{bmatrix} =
 \begin{bmatrix}
     i & 0 & 0 & -i\mathrm{e}^{-i\theta_k} \\
     0 & -i & i\mathrm{e}^{i\theta_k} & 0 \\
     0 & 0 & -i & 0 \\
     0 & 0 & 0 & i \\
 \end{bmatrix}
  \begin{bmatrix}
    \hat{c}_k^\dagger \\ \hat{h}_k^\dagger \\ \hat{c}_k \\ \hat{h}_k
 \end{bmatrix} \;\;
\end{equation}
which  encompasses rotations, translations, scaling, and shearing. The inverse gives
\begin{equation}
 \begin{bmatrix}
    \hat{c}_k^\dagger \\ \hat{h}_k^\dagger \\ \hat{c}_k \\ \hat{h}_k
 \end{bmatrix} =
 \begin{bmatrix}
     -i & 0 & 0 & -i\mathrm{e}^{-i\theta_k} \\
     0 & i & i\mathrm{e}^{i\theta_k} & 0 \\
     0 & 0 & i & 0 \\
     0 & 0 & 0 & -i \\
 \end{bmatrix}
  \begin{bmatrix}
    \hat{a}_k^\dagger \\ \hat{b}_k^\dagger \\ \hat{a}_k \\ \hat{b}_k
 \end{bmatrix} \;\;.
\end{equation}
It can be calculated that
\begin{equation}
    [\hat{h}_k^\dagger,\hat{c}_k^\dagger] = 2i \sin{\theta_k} \;\;,
\end{equation}
and all other commutators vanish. This indicates that the transformation has distorted the extended space leading to a more complex structure. In particular, operators $\hat{c}_k^\dagger$ as well as $\hat{h}_k^\dagger$ must be understood to be no longer independent so that it may be difficult to extract the physical information from the ADOs. 
On the other hand, it can be seen that number operators obey
\begin{equation}
    [\hat{c}^\dagger\hat{c}, \hat{h}^\dagger\hat{h}] = 2i\sin{\theta_k} \hat{a} \hat{b} \;\; .
\end{equation}

By applying the above transformation to Eq. (\ref{Eq:standard-heom-liouville-fock}), a more symmetric form is achieved, i.e.,
\begin{multline}
\label{Eq:HEOM-lindblad}
    \dot{\hat{W}}
= -i\mathcal{L}_\mathrm{s} \hat{W}  + \sum_{k=1}^{K}
\left\{ [2\gamma_k^{\mathrm{eff}}\hat{c}_k\hat{h}_k -z_k\hat{c}_k^\dagger\hat{c}_k - z_k^\ast\hat{h}_k^\dagger\hat{h}_k] \hat{W} \right. 
\\ \left. + i\sqrt{d_k}\hat{q}_s (\hat{c}_k+\hat{c}_k^\dagger) \hat{W} - i\sqrt{d_k^\ast} (\hat{h}_k +\hat{h}_k^\dagger) \hat{W}\hat{q}_s 
 \right\} \;\;.
\end{multline}
In the above, the effective damping $\gamma_k^{\rm eff}$ is defined through
\begin{equation}\label{Eq:eff_damping}
    \gamma_k^{\rm eff} = \gamma_k\cos{\theta_k} + \omega_k\sin{\theta_k}\;\;.
\end{equation}

In parallel to Eq.~(\ref{Eq:standard-heom-density}), this expression can be mapped onto a time evolution equation in Lindblad-like form for the density matrix, i.e., 
\begin{multline}
    \dot{\hat{\rho}} = -i [\hat{H}_{\rm eff}, \hat{\rho} ] + \sum_{k=1}^K[ 2\gamma_k^{\rm eff} \hat{c}_k  \hat{\rho}\hat{c}_k^{\dagger} -\gamma_k \{\hat{c}_k ^{\dagger} \hat{c}_k, \hat{\rho} \}\;\;,
\end{multline}
with a non-Hermitian Hamiltonian defined as
\begin{equation}
    \hat{H}_{\rm eff} = \hat{H}_s + \hbar\omega_k\hat{c}_k^\dagger\hat{c}_k - \sqrt{d_k}\hat{q}_s (\hat{c}_k^\dagger + \hat{c}_k) \;\;.
\end{equation}
 In comparison to the regular Lindblad structure, damping is captured by an effective damping $\gamma_k^{\rm eff}$ instead of $\gamma_k$. 

\section{From the Stochastic Liouville-von Neumann Equation to the FP-HEOM}

We here provide an alternative way to connect the FP-HEOM with the the Stochastic Liouville-von Neumann Equation (SLN), where one starts from the latter and goes all the way back to the FP-HEOM.

The SLN reads\cite{stockburger02} 
\begin{equation}\label{Eq:sln}
    \frac{d}{dt} \hat{\rho}_\mu = -i \hat{\mathcal{L}}_s\hat{\rho}_\mu +i \xi(t) [\hat{q}_\mathrm s, \hat{\rho}_\mu] + i \frac{\nu(t)}{2} \{ \hat{q}_\mathrm s, \hat{\rho}_\mu \} \;\;
\end{equation}
and evolves density matrices $\hat{\rho}_\mu$ depending on a set of stochastic fields $\mu=\{\xi, \nu\}$ which obey $\mathcal{M}_\mu[\xi(t)\xi(v) ]= {\rm Re}\, C(t - v)$, $\mathcal{M}_\mu[\xi(t)\nu(v)] = 2i\Theta(t-v){\rm Im}\,C(t-v) $ with $\Theta$ being step function, and $\mathcal{M}_\mu[\nu(t)\nu(v) ]= 0$. The physical density matrix follows from $\hat{\rho}_\mathrm s=\mathcal{M}_\mu\{\hat{\rho}_\mu\}$, where $\mathcal{M}_\mu$ denotes the mean value over the set of stochastic fields. 

Now, in order bring this stochastic time evolution into contact with the time evolution in Liouville-Fock-space according to eq.\eqref{Eq:standard-heom-liouville-fock}, we introduce two functional differential operators with regard to the stochastic fields, i.e., 
\begin{equation}
\begin{aligned}
    \hat{\phi}_k(t) &= \frac{1}{2i} \int\limits_{-\infty}^{+\infty} ds  \sqrt{d_k} e^{-z_k(t-s)} \left( \frac{\delta}{\delta \xi(s)} + 2 \frac{\delta}{\delta \nu(s)} \right) \;\;  \\
    \hat{\psi}_k(t) &= \frac{i}{2} \int\limits_{-\infty}^{+\infty} ds  \sqrt{d_k^*} e^{-z_k^*(t-s)} \left( \frac{\delta}{\delta \xi(s)} - 2 \frac{\delta}{\delta \nu(s)} \right)\; . 
\end{aligned}
\end{equation}
These functionals satisfy the relations $\partial_t \hat{\phi}_k = -z_k\hat{\phi}_k$ and $\partial_t \hat{\psi}_k = -z_k^*\hat{\psi}_k$. This allows us to define a \textit{stochastic} extended operator in Liouville-Fock-space via
\begin{equation}
    \hat{W}_\mu(t) = \sum_{{\bf m,n}} \prod_{k = 1}^K \frac{1}{\sqrt{m_k! n_k!}} \hat{\phi}_k^{m_k}(t) \hat{\psi}_k^{n_k}(t) \hat{\rho}_\mu(t) | {\bf m,n} \rangle \; ,
\end{equation}
such that the reduced density matrix is the lowest order expansion coefficient after taking the average over the stochastic fields $\hat{\rho}_s = \langle {\bf 0,0}| \mathcal{M}_\mu[\hat{W}_\mu]$. Furthermore, by using the the occupation operators $\hat{m}_k = \hat{a}_k^\dagger \hat{a}_k$ and $\hat{n}_k = \hat{b}_k^\dagger \hat{b}_k$, it is easily verified that $\hat{\phi}_j \hat{W}_\mu = \hat{a}_j \hat{W}_\mu$ and $\hat{\psi}_j \hat{W}_\mu = \hat{b}_j \hat{W}_\mu$ holds and lets us simplify the Novikov theorem 
\begin{equation}
\begin{aligned}
    \mathcal{M}_\mu[\xi(t)\hat{W}_\mu(t)] &= \int\limits_{-\infty}^\infty ds \mathcal{M}_\mu[\xi(t)\xi(s)] \mathcal{M}_\mu\left[\frac{\delta \hat{W}_\mu(t)}{\delta \xi(s)}\right] \\
    &+ \int\limits_{-\infty}^\infty ds \mathcal{M}_\mu[\xi(t)\nu(s)] \mathcal{M}_\mu\left[\frac{\delta \hat{W}_\mu(t)}{\delta \nu(s)}\right] \\
    &= i\sum_{j = 1}^K\left(  \sqrt{d_j} \hat{a}_j\mathcal{M}_\mu[\hat{W}_\mu(t)] -  \sqrt{d_j^*}\hat{b}_j\mathcal{M}_\mu[\hat{W}_\mu(t)]\right)\label{Eq:sln_novikov}
\end{aligned}
\end{equation}
als well as $\mathcal{M}_\mu[\nu(t)\hat{W}_\mu(t)] = 0$. The time evolution of eq.\eqref{Eq:sln} can be recast into  
\begin{multline}
    \frac{d}{dt}\hat{W}_\mu(t) = -i\hat{\mathcal{L}}_S \hat{W}_\mu(t) - \sum_{j = 1}^K \left(\hat{a}_j^\dagger\hat{a}_j z_j  + \hat{b}_j^\dagger \hat{b}_j z_j\right)\hat{W}_\mu(t)\\
    + i \sum_{{\bf m,n}} \prod_{k=1}^K \frac{\hat{\phi}_k^{m_k}(t)\hat{\psi}_k^{n_k}(t)}{\sqrt{m_k!n_k!}}\xi(t) [\hat{q}_s,\hat{\rho}_\mu(t)] |{\bf m,n} \rangle\\
    +\frac{i}{2} \sum_{{\bf m,n}} \prod_{k=1}^K \frac{\hat{\phi}_k^{m_k}(t)\hat{\psi}_k^{n_k}(t)}{\sqrt{m_k!n_k!}}\nu(t) \{\hat{q}_s,\hat{\rho}_\mu(t)\} |{\bf m,n} \rangle \; .
\end{multline}
In order to proceed further we will need to commute the stochastic fields with the functional differential operator. The commutation relation at time $t$ for this operation is 
\begin{multline}\label{Eq:sln_commutator}
    \left[ \prod_{j=1}^K \frac{\hat{\phi}_j^{m_j}\hat{\psi}_j^{n_j}}{\sqrt{m_j!n_j!}},\xi\right]\hat{f} = \sum_{j = 1}^K \Biggl( \frac{\sqrt{d_j}}{i} \frac{\hat{\phi}_j^{m_j-1}\hat{\psi}_j^{n_j}}{\sqrt{(m_j - 1)!n_j!}}\sqrt{m_j}\\
    + \frac{i \sqrt{d_j^*}}{2}\frac{\hat{\phi}_j^{m_j}\hat{\psi}_j^{n_j-1}}{\sqrt{m_j!(n_j-1)!}}\sqrt{n_j} \Biggl)\prod_{l\neq j}\frac{\hat{\phi}_l^{m_l}\hat{\psi}_l^{n_l}}{\sqrt{m_l!n_l!}} \hat{f}\; .
\end{multline}
The commutator $[\prod_k \hat{\phi}_j^{m_j}\hat{\psi}_j^{n_j}/\sqrt{m_j!n_j!},\nu]$ is eq.\eqref{Eq:sln_commutator} where one replaces $\sqrt{d_j}/2i \rightarrow \sqrt{d_j}/i$ and $i\sqrt{d_j^*}/2 \rightarrow -i\sqrt{d_j^*}$. We thus arrive at 
\begin{multline}
    \frac{d}{dt} \hat{W}_\mu(t) = - i\hat{\mathcal{L}}_\mathrm s\hat{W}_\mu(t) - \sum_{j = 1}^K \left(\hat{a}_j^\dagger\hat{a}_j z_j  + \hat{b}_j^\dagger \hat{b}_j z_j  \right)\hat{W}_\mu(t)\\
    + i \xi(t) [\hat{q}_s,\hat{W}_\mu(t)] + i\frac{\nu(t)}{2}\{ \hat{q}_\mathrm s,\hat{W}_\mu(t) \} \\
    + \sum_{j=1}^K \left( \sqrt{d_j}\hat{q}_\mathrm s\hat{a}_j^\dagger\hat{W}_\mu(t) + \sqrt{d_j^*}\hat{b}_j^\dagger\hat{W}_\mu(t)\hat{q}_\mathrm s \right)\; .
\end{multline}
Taking the expectation value on both sides of the equation above and defining the deterministic extended operator $\hat{W} = \mathcal{M}_\mu[\hat{W}_\mu]$ in Liouville-Fock-space will yield Eq. \eqref{Eq:standard-heom-liouville-fock} after using the Novikov theorem in eq.\eqref{Eq:sln_novikov}, \textit{i.e.} 
\begin{multline}
    \frac{d}{dt} \hat{W} = - i\hat{\mathcal{L}}_\mathrm s\hat{W} - \sum_{j = 1}^K \left(\hat{a}_j^\dagger\hat{a}_j z_j  + \hat{b}_j^\dagger \hat{b}_j z_j  \right)\hat{W} \\
    - \sum_{j = 1}^K \left[\hat{q}_s, \sqrt{d_j}\hat{a}_j\hat{W} - \sqrt{d_j^*}\hat{b}_j\hat{W}\right]\\
    + \sum_{j=1}^K \left( \sqrt{d_j}\hat{q}_\mathrm s\hat{a}_j^\dagger\hat{W} + \sqrt{d_j^*}\hat{b}_j^\dagger\hat{W}\hat{q}_\mathrm s \right)
\end{multline}

\bibliography{quantum}

\begin{thebibliography}{141}
\bibitem [{mar()}]{mark1}%
  \BibitemOpen
  \href@noop {} {\bibinfo {title} {{We note that in the most general scenarios,
  which include very low temperatures and a broad class of reservoir spectral
  densities such as those that are highly structured, complex-valued
  frequencies $z_k$ and amplitudes $d_k$ offer accelerated convergence. In
  specific situations, like those with a smooth noise spectrum
  $S_\beta(\omega)$, either of these two complex-valued parameters may reduce
  to real or imaginary values.}}}\BibitemShut {Stop}%
\end{thebibliography}
\end{document}